\documentclass[12pt]{article} 
\pdfoutput=1
\usepackage{amssymb,graphicx}
\usepackage{epstopdf}
\usepackage{amsmath,amsfonts}
\usepackage{epsfig} 
\usepackage{graphicx,graphics}
\usepackage[dvipsnames]{xcolor}

\usepackage[unicode=true,pdfusetitle,
bookmarks=true,bookmarksnumbered=true,bookmarksopen=true,bookmarksopenlevel=3,
 breaklinks=false,pdfborder={0 0 1},backref=false,colorlinks=true,citecolor=blue]
 {hyperref}

\begin{document} 
\begin{flushright}
	INR-TH-2025-017
\end{flushright}

\title{\bf Cosmic domain walls on a lattice:\\ illusive effects of initial conditions}
\author{I.~Dankovsky$^{a,b}$, S.~Ramazanov$^c$, E.~Babichev$^d$, D.~Gorbunov$^{b,e}$, A.~Vikman$^f$\\
\small{\em $^a$Faculty of Physics, Lomonosov Moscow State University, 119991 Moscow, Russia}\\
\small{\em $^b$Institute for Nuclear Research of the Russian Academy of Sciences, 117312 Moscow, Russia}\\
\small{\em  $^c$Institute for Theoretical and Mathematical Physics, MSU, 119991 Moscow, Russia}\\ 
\small{\em  $^d$Universit\'e Paris-Saclay, CNRS/IN2P3, IJCLab, 91405 Orsay, France}\\ 
\small{\em  $^e$Moscow Institute of Physics and Technology, 141700 Dolgoprudny, Russia}\\ 
\small{\em  $^f$CEICO, FZU-Institute of Physics of Czech Academy of Sciences,}\\
\small{\em Na Slovance 1999/2, 182 00 Prague, Czech Republic}}

 \date{}

{\let\newpage\relax\maketitle}

\begin{abstract}
Evolution of cosmic domain walls (DWs) settles to the scaling solution, which is often assumed to be independent of initial conditions. However, lattice simulations performed in this work reveal a clear dependence of the scaling DW area on the initial configuration of the sourcing scalar field, specifically, its infrared (IR) properties. Namely, the DW area grows as one suppresses IR modes in the initial scalar field spectrum. This growth is saturated, when the area parameter $\xi$ commonly used in the literature reaches the value $\xi_{max} \approx 1.2$. The dependence of $\xi$ on IR modes is argued to be of non-physical origin: it is likely to be due to effects of the lattice boundary. Assuming that {\it physically} the memory of initial conditions is erased, one recognizes $\xi \approx 1.2$ obtained in the situation with maximally suppressed IR modes as a genuine universal value of the area parameter in the scaling regime. We demonstrate that ignorance about initial conditions may affect predictions for the energy density of gravitational waves by the factor ${\cal O} (5)$. The spectral shape of gravitational waves is also affected by the choice of initial conditions, most notably in the low-frequency part.
Likewise, we revisit annihilation of DWs under the influence of a potential bias. It has been previously found in Ref.~\cite{bias} 
that the annihilation happens significantly earlier compared to the estimate based on the simple balance between the potential bias and surface energy density. We further support this observation and show that the tendency towards an earlier annihilation gets even stronger upon removing IR modes in simulations.

\end{abstract}

\section{Introduction} 
\label{sec:intro}

Formation of cosmic domain walls (DWs) is common in particle physics scenarios involving spontaneously broken discrete symmetries~\cite{Zeldovich:1974uw}, e.g., in axion models~\cite{Sikivie:1982qv}. Unless one chooses their tension to be very small, DWs carry a large energy density causing troubles for the cosmological concordance model. To avoid a conflict with observations, one assumes that the DW network gets destroyed in one or another way in the early Universe~\cite{Gelmini:1988sf, Lazarides:1982tw, Coulson:1995nv, Vilenkin:1981zs, Ramazanov:2021eya, Babichev:2021uvl, Babichev:2023pbf}. 
Nevertheless DWs can leave profound traces on cosmological observables through particle~\cite{Hiramatsu:2012gg, Kawasaki:2014sqa} or gravitational wave (GW) emission~\cite{Hiramatsu:2013qaa, Ferreira:2022zzo, Ferreira:2023jbu, Kitajima:2023cek, first, Cyr:2025nzf, Notari:2025kqq, bias}. Lattice simulations provide with an important tool for studying DW networks and their phenomenological consequences. In this way, many properties of DWs have been uncovered, but it would be fair to say that the DW science is still in its infancy and many questions remain open. In this paper, we investigate a subtle issue of sensitivity of DW network dynamics to the statistical properties of initial conditions of the scalar field 
that forms the 
network. 

A DW network enjoys a nice property of settling to the scaling regime soon after formation~\cite{Hiramatsu:2013qaa, Press:1989yh, Garagounis:2002kt}. 
That is, the long range dynamics of DW network proceeds in the self-similar manner with the characteristic length scale given by the Hubble radius. The scaling solution describing DW evolution 
is assumed to be an attractor reached independently of initial conditions. This statement, however, has not been tested 
in the literature. In fact, simulations of Refs.~\cite{Ferreira:2023jbu, first} suggest that different types of initial conditions may end up with different solutions 
for the late time DW network, cf. Refs.~\cite{Kitajima:2023kzu, Gonzalez:2022mcx}. These differences somewhat affect predictability of scenarios 
involving DWs. It is the goal of this work to understand, if such a dependence is physical or due to some artifact of simulations.

Continuing the line of research of Ref.~\cite{first}, we are collecting more evidence that lattice evolution of DWs depends on initial conditions. For this purpose we use the  parameter
\begin{equation}
\label{area}
\xi(t) =\frac{S \cdot t}{a(t) V} \;,  
\end{equation}
where $t$ is the cosmic time, $a(t)$ is the scale factor of the expanding Universe, while $S$ is the comoving DW area captured at this moment of time within the comoving volume $V$. We use the terms scaling parameter and area parameter interchangeably for the quantity $\xi$. The latter takes a constant value in the scaling regime, and we observe such a behavior for the majority of utilized  initial conditions. Nevertheless, this constant varies as a function of initial conditions, 
so that the scaling parameter takes values in the range $0.5 \lesssim \xi \lesssim 1.2$. We demonstrate 
that this dependence is mainly due to IR properties of initial conditions. Namely, suppressing IR modes in the initial scalar power spectrum, one increases the scaling parameter $\xi$ until it reaches the maximal value $\xi_{max} \approx 1.2$. We argue, however, that such a dependence on IR modes is likely to be of non-physical origin, and it is caused by the lattice boundary effects, 
to which long wavelength scalar field fluctuations are particularly sensitive. To avoid such a dependence in the numerical simulations, one should carefully select initial conditions by properly removing/suppressing  
long wavelength modes\footnote{Note that boundary effects inevitably affect lattice simulations at late times, 
because the simulated Hubble patches grow faster than the box size, 
and one may end up with no DWs inside the lattice box. 
To avoid this situation, one limits duration of simulations, which is also required to be able to resolve the DW width. In this work, we discuss another way of how boundary effects enter DW evolution: through the IR modes.}. We extrapolate the value $\xi \approx 1.2$ resulting in this case to arbitrary initial conditions. This is justified, provided that physically DWs carry no dependence on initial conditions in the scaling regime.

Likewise, we study the impact of initial conditions on lattice simulations of DW annihilation in sec.~\ref{sec:ann}. Recall that DWs are very heavy and tend to overclose the Universe, so that their annihilation must take place at some moment of the cosmic evolution. A common way to destroy the DW network is by introducing a potential bias $V_{bias}$~\cite{Zeldovich:1974uw, Gelmini:1988sf}, which explicitly breaks the $Z_2$-symmetry. Typically one assumes that the annihilation of DWs occurs at the time $t_{ann} \sim \sigma_{wall}/V_{bias}$~\cite{Vilenkin:1981zs}, where $\sigma_{wall}$ is the DW tension. In Ref.~\cite{bias}, however, it has been observed that the behavior $t_{ann} \propto V^{-2/3}_{bias}$ agrees better with the simulation data. Thus, the annihilation may happen much earlier than naively expected, since one usually deals with a very small $V_{bias}$. We check in sec.~\ref{sec:ann}, if this conclusion is robust against modifying the initial scalar field configuration. We find that the tension with a commonly assumed relation $t_{ann} \propto V^{-1}_{bias}$ becomes more pronounced upon removing IR modes. The dependence $t_{ann} \propto V^{-2/3}_{bias}$ still provides a good fit to the simulation data, but some datasets show a clear preference for the behavior $t_{ann} \propto V^{-1/2}_{bias}$.

Finally, in sec.~\ref{sec:gw} we investigate the impact of initial conditions 
on the lattice predictions for GWs, produced by the DW network. This research is primarily motivated by the recently observed GW background in the nanohertz frequency range in the PTA data~\cite{NANOGrav:2023gor, EPTA:2023fyk, EPTA:2023xxk, Xu:2023wog, Reardon:2023gzh}. DWs are among promising candidates for the source of the GW background found~\cite{NANOGrav:2023hvm}, 
and they deserve a thorough analysis in this regard. With the data from the currently planned space-born LISA~\cite{LISA:2017pwj}, Taiji~\cite{Hu:2017mde,Ruan:2018tsw}, TianQin~\cite{TianQin:2015yph}, DECIGO~\cite{Kawamura:2020pcg} and the ground-based Einstein Telescope (ET)~\cite{Abac:2025saz} and Cosmic Explorer~\cite{Reitze:2019iox} one will also be able to probe DW models predicting GWs in frequency ranges many orders of magnitude above the PTA one. In sec.~\ref{sec:gw} we revisit both cases of GW emission from unbiased and biased DWs. We observe that variation of initial conditions can affect predicted intensity of GWs by the factor ${\cal O} (5)$. The generated GW spectra 
have less power in the long wavelength regime in the situation with suppressed IR modes in the initial scalar distribution, --- this feature is particularly prominent in the case of biased DWs. 
We also confirm existence of a plateau-like feature in the high frequency part of the GW spectrum observed earlier 
in Refs.~\cite{Ferreira:2023jbu, Kitajima:2023cek, first}.

\section{The model}
\label{sec:model}

We consider the model of a real scalar field $\chi$ with the canonical kinetic term and the potential exhibiting spontaneous breaking of an approximate $Z_2$-symmetry:
\begin{equation}
\label{action}
{\cal L}=\frac{1}{2} \left(\partial \chi \right)^2-\frac{\lambda}{4} (\chi^2-v^2)^2-\epsilon \chi^3 \; ,
\end{equation}
where $\lambda$ and $v$ are the self-interaction coupling constant and the expectation value of the scalar $\chi$.
The last term on the r.h.s. of Eq.~\eqref{action}, controlled by the dimension-of-mass parameter $\epsilon$, is responsible for an explicit breaking of $Z_2$-symmetry. 
Such a term introduces an asymmetry in the potential minima, and it is responsible for the annihilation of a DW network. In this work we consider two different  situations: one with $\epsilon =0$, in which case the DW network is eternal, and one with $\epsilon \neq 0$, in which case it gets destroyed at some point. We refer to DWs as unbiased and biased in these two cases, respectively. 
We assume that DWs form and evolve during the radiation dominated (RD) stage. In the RD era the scale factor grows with conformal time $\tau$ as $a(\tau) \propto \tau$, and the Hubble rate decreases as $H =1/(a \cdot \tau) \propto 1/\tau^2$. Neglecting metric perturbations, 
one writes the equation of motion for the scalar $\chi$:
\begin{equation}
\label{eq}
\chi''+\frac{2}{\tau} \chi'-\partial_i \partial_i \chi + \lambda \chi a^2 (\tau)(\chi^2-v^2)+3\epsilon a^2 (\tau) \chi^2=0 \; ,
\end{equation}
where the prime $'$ denotes the derivative with respect to the conformal time $\tau$. Hereafter we switch to dimensionless variables: $\chi \rightarrow \chi/v$, $\tau \rightarrow \sqrt{\lambda} v \tau$, $x^i \rightarrow \sqrt{\lambda} v x^i$, and $\epsilon \rightarrow \epsilon/(\lambda v)$. We also choose the moment when $H_i=\sqrt{\lambda}v$ as the starting point of the network evolution. Furthermore, fixing $\tau_i=1$, one can write the equation of motion~\eqref{eq} in the form:
\begin{equation}
\label{eqdim}
\chi''+\frac{2}{\tau} \chi'-\partial_i \partial_i \chi + \chi \tau^2 (\chi^2-1)+3\epsilon \tau^2 \chi^2=0 \; .
\end{equation}
Here we inserted $a_i=1$, which follows from $H_i=1/(a_i \tau_i)= \sqrt{\lambda} v$ and $\tau_i=1/(\sqrt{\lambda} v)$ (in the original dimensionful units).

The DW network forms as the scalar $\chi$ rolls toward the minima of its potential.
Away from a short initial period, evolution of DWs proceeds in the self-similar manner. In this regime, DWs are sufficiently smooth with a network correlation length being of the order of the horizon size. Another characteristic length scale in the problem ---DW width
\begin{equation}
\label{eq:width}
\delta_{wall} =\sqrt{\frac{2}{\lambda}}\,\frac{1}{v}\,,
\end{equation}
quickly becomes much smaller than the horizon radius and it is not manifested in the large scale dynamics of DWs. Such thin DWs are 
described by a tension or surface energy density $\sigma_{wall}$ given by 
\begin{equation}
\sigma_{wall} =\frac{2\sqrt{2\lambda} v^3}{3} \; .
\end{equation}
The tension is subjected to stringent restrictions $\sigma_{wall} \lesssim (1~\mbox{MeV})^3$ in the case of unbiased DW~\cite{Zeldovich:1974uw, Lazanu:2015fua}. The bound is avoided in the case $\epsilon \neq 0$, i.e. for the biased DWs.

As it follows, neither $\lambda$, nor $v$ enters the dimensionless equation of motion~\eqref{eqdim}, and therefore they can be set to arbitrary values, 
when running numerical simulations. However, these parameters are relevant when defining the fractional energy density of DWs in the Universe and the amplitude of GWs. We set them to $\lambda=0.03$ and $v=6 \cdot 10^{16}$\,GeV. The 
Hubble rate at the onset of simulations is fixed by the condition $H_i=\sqrt{\lambda} v$, which gives $H_i \approx 10^{16}$\,GeV. The fact that it is unrealistically large is not important. By proper rescaling
of the simulation results, one can always adjust them to the interesting values of the model parameters $\lambda$ and $v$. 
One should only ensure that the energy density 
of DWs does not surpass that of the surrounding energy density of radiation, as we want to keep DW component  only subdominant in the expanding  Universe. 
DWs become dominant, when $\sigma_{wall} H \approx 3H^2 M^2_{P}$, where $M_{P}$ is the reduced Planck mass\footnote{We use units $c=\hbar=k_{B}=1$, so that $M_{P}=2.435\times10^{18}$~GeV.}. This gives for the corresponding conformal time $\tau_{dom} \approx 70$. As we discuss in sec.~\ref{sec:setting}, there are more severe constraints on the time range of simulations from a finite lattice size and lattice spacing. 

We set up initial conditions for the Fourier modes of the scalar $\chi$, 
\begin{equation}
\chi ({\bf k}) = \int d{\bf x} e^{-i {\bf kx}} \chi ({\bf x}) \; ,
\end{equation}
by introducing two-point functions: 
\begin{equation}
\label{init}
\langle \chi ({\bf k}) \chi ({\bf q}) \rangle =(2\pi)^3 A(k) \delta ({\bf k}+{\bf q})\,, \qquad \langle \chi' ({\bf k}) \chi' ({\bf q}) \rangle =(2\pi)^3 B(k) \delta ({\bf k}+{\bf q}) \; .
\end{equation}
To demonstrate the main statement, we choose the following family of artificial initial conditions: 
\begin{equation}
\label{AB}
A(k)=\frac{\mbox{exp} \left(-\frac{k^2_{IR}}{k^2} \right) \cdot \theta \left(k_{UV}-k \right)}{k \left(e^{\frac{k}{T}}-1 \right)^{\alpha}}\,, \qquad B(k)=\frac{k \cdot \mbox{exp} \left(-\frac{k^2_{IR}}{k^2}\right) \cdot \theta \left(k_{UV}-k \right)} {\left(e^{\frac{k}{T}}-1 \right)^{\alpha}}  \; .
\end{equation}
By varying the free parameter $\alpha$, as well as cutoff scales $k_{IR}$ and $k_{UV}$, we can manipulate the initial conditions and study their impact on DW evolution. 
Modulo the factor $2$, the case $\alpha=0$ corresponds to vacuum initial conditions for a massless scalar in the flat spacetime, and it is the one most commonly adopted in the literature. The case $\alpha=1$ corresponds to the thermal initial conditions. These cases have been studied in Ref.~\cite{first} and we found clear differences in DW evolution. One of the goals of the present work is to demonstrate if such differences are generic by considering a larger scope of parameters $\alpha$. The temperature is taken to be $T=1/5 T_{U} (\tau_i)$, where $T_{U}$ is the temperature of the primordial plasma, which is related to the Hubble rate in the RD epoch as 
$H=\sqrt{\pi^2 g_* (T_U)/90} \cdot T^2_{U}/M_{P}$, where $g_* (T_U)$ is the number of relativistic degrees of freedom, which we fix to be $g_* (T_U) =100$. The reason for such a choice of $T$ is technical: for high temperatures $T$ initial fluctuations of the field $\chi$ are also quite large. This may cause problems when counting the DW area, i.e., inhomogeneities of the field $\chi$ can be misinterpreted as a DW. Such issues do not arise at the lower temperatures used in our study. Note that with our choice of parameters one has $T \approx 1$.

Cutoff scales $k_{IR}$ and $k_{UV}$ in Eq.~\eqref{AB} bound the scalar field spectrum from below and from above, respectively. The UV cutoff $k_{UV}$ is imposed by means of the theta-function $\theta (k_{UV}-k)$, which equals unity for $k \leq k_{UV}$ and vanishes otherwise. The reasons, why we apply the IR and UV cutoff, are twofold. First, in the case $\alpha \leq 0$, the scalar $\chi$ variance is unbounded, 
unless one chooses a finite UV cutoff. More crucially, however, one would like to test sensitivity of DW evolution 
upon the inclusion/elimination of IR or UV modes. Here let us also elucidate some of the terminology to be used in what follows. We refer to modes with $k \lesssim 1$, which are superhorizon at the time $\tau_i=1$, and $k \gtrsim 1$, which are subhorizon at $\tau_i=1$, as IR and UV ones, respectively. Recall that we are working with dimensionless variables, and the momenta $k$ are in units of $\sqrt{\lambda} v$. Note that the parameter $\alpha$ in Eq.~\eqref{AB} defines the slope of deeply IR modes in the situation with $k_{IR}=0$. Indeed, focusing on $k \ll 1$, one can write 
\begin{equation}
\label{alphair}
\langle \chi^2 ({\bf x}) \rangle_{k \ll 1} =\int \frac{dk}{k} \frac{k^{2-\alpha} T^{\alpha}}{2\pi^2} \; .
\end{equation}
As it follows, one distinguishes the cases $\alpha \geq 2$ and $\alpha<2$, for which the scalar $\chi$ spectrum is dominated by IR modes and UV modes, respectively.

\section{Setting up a lattice}
\label{sec:setting}

All the numerical simulations discussed throughout this paper have been carried out on a lattice using the public code CosmoLattice~\cite{Figueroa:2020rrl, Figueroa:2021yhd}. 
The lattice box characterized by the comoving size $L$ mimics the expanding Universe, which contains many causally disconnected Hubble patches, at least initially. In what follows, we use lattices with $N^3=1024^3$ and $N^3=2048^3$ grid points. 
Periodic conditions are imposed on the lattice boundaries. Due to a finite box size, the lowest comoving momentum characterizing Fourier modes of the field $\chi$ is given by
\begin{equation}
k_{min}=\frac{2\pi}{L} \; . 
\end{equation}
This also plays the role of an effective IR cutoff for initial fluctuations when $k_{IR}=0$ in Eq.~\eqref{AB}.

When running simulations, one needs to ensure that the lattice spacing at the end of simulations is not exceeding the DW width $\delta_{wall} \simeq \sqrt{2}$ 
(recall that all the distances are measured in units of $(\sqrt{\lambda} v)^{-1}$). That is, introducing the parameter $\kappa$ defined as the ratio of the DW width $\delta_{wall}$ and the final lattice spacing, 
\begin{equation}
\label{kappa}
\kappa \equiv \frac{\delta_{wall} N}{a(\tau_f) L} \; ,
\end{equation}
one requires that it is not smaller than unity, $\kappa \gtrsim 1$. Another important condition to be fulfilled is that the lattice size is twice larger than the Hubble radius by the end of simulations. Namely, one requires that the parameter $\kappa'$, defined as  
\begin{equation}
\label{kappaprime}
\kappa' \equiv \frac{L}{2 \tau_f}\,,
\end{equation}
is also not smaller than unity, $\kappa' \gtrsim 1$. Using Eqs.~\eqref{kappa},~\eqref{kappaprime}, and 
$H_i =\sqrt{\lambda} v$, one obtains the time span of simulations in terms of parameters $\kappa$ and $\kappa'$:
\begin{equation}
\label{timespan}
\tau_f \approx \frac{\sqrt{N}}{2^{1/4} \sqrt{\kappa \kappa'}} \; .
\end{equation}
One can also express the optimal lattice comoving size through the grid number $N$ and parameters $\kappa$ and $\kappa'$: 
\begin{equation}
\label{optimal}
L =2^{3/4} \sqrt{\frac{N \kappa'}{\kappa}} \; .
\end{equation}
Although the choice $\kappa'=\kappa=1$ is the optimal one, as it ensures the longest time interval of simulations and minimization of effects caused by the lattice box, in what follows we allow $\kappa'$ to deviate slightly from unity. That is, we mainly assume $\kappa'=\pi/2$ and $\kappa=1$ throughout the paper, and that Eq.~\eqref{optimal} is fulfilled. 
As it follows from Eq.~\eqref{kappaprime}, the lattice box contains approximately $(L/\tau_f)^3 \approx 30$ Hubble volumes at the time $\tau_f$ for this choice of $\kappa'$.
In the case of $1024^3$ and $2048^3$ lattices, Eq.~\eqref{timespan} gives for the final time of simulations $\tau_f \approx 22$ and $\tau_f \approx 30$, respectively. Note, however, that we assume a departure from Eq.~\eqref{optimal} in sec.\,\ref{sec:impact} where studying lattice boundary effects, and varying $\kappa'/\kappa$ in sec.\,\ref{sec:gw} where studying GWs. We comment on the corresponding limitations on the final time $\tau_f$, where relevant.

\section{Impact of initial conditions on DW evolution:\\ unbiased case}
\label{sec:impact}

\begin{figure}[!htb]
\begin{center}
    \includegraphics[width=\textwidth]{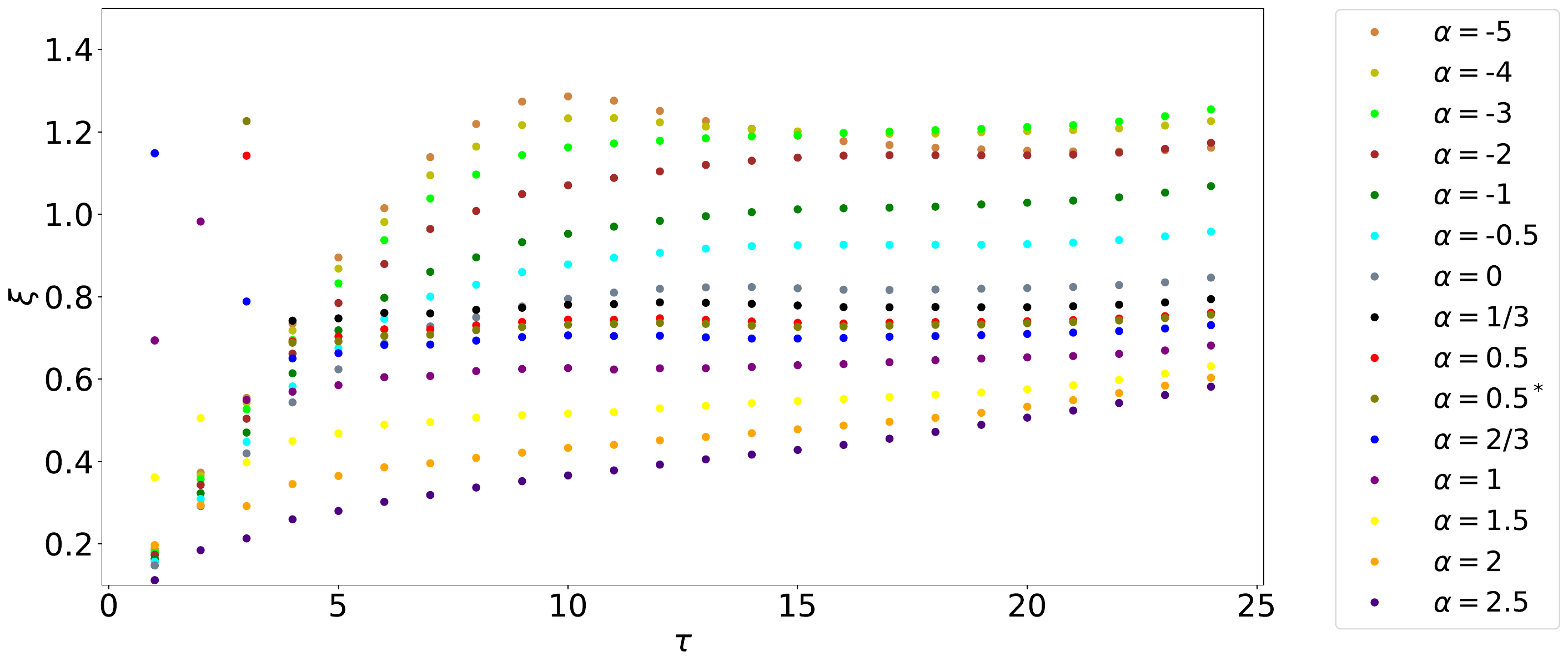} 
\end{center}
    \caption{Evolution of the area parameter $\xi$ for different choices of the parameter $\alpha$ characterizing initial conditions in Eq.~\eqref{AB}. An exception is the case $\alpha=0.5$ marked with an asterisk, in which case the r.h.s. of Eq.~\eqref{AB} is multiplied by a factor of $2$, i.e., we have assumed doubled power of the initial scalar spectrum. The IR cutoff is set to $k_{IR}=0$. No UV cutoff is set for $\alpha>0$, while for $\alpha \leq 0$ it is set to $k_{UV}=1$. Simulations have been carried out using $1024^3$ lattice with the box size $L$ given by Eq.~\eqref{optimal}.} \label{scaling}
\end{figure}

\begin{figure}[!htb]
\begin{center}
    \includegraphics[width=0.9\textwidth]{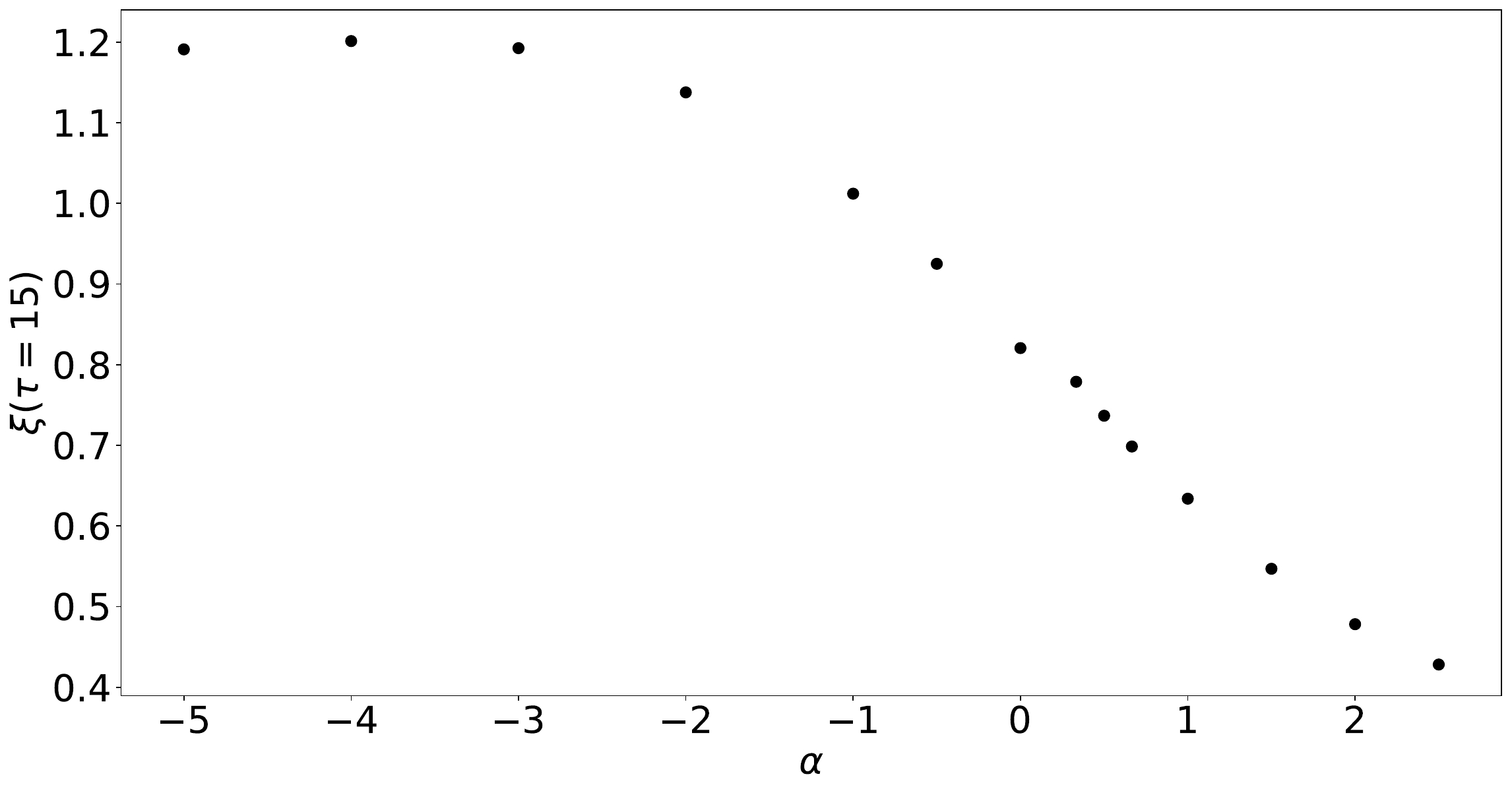} 
\end{center}
    \caption{Dependence of the area parameter $\xi$ taken at the conformal time $\tau=15$ on the parameter $\alpha$ characterizing initial conditions in Eq.~\eqref{AB}. The dependence has been inferred from Fig.~\ref{scaling}.} \label{alpha}
\end{figure}

We start numerical simulations with the evolution of unbiased DWs using the initial conditions described by Eqs.~\eqref{init} and~\eqref{AB}. In Fig.\,\ref{scaling}  
we demonstrate the dependence of the area parameter $\xi$ defined by Eq.~\eqref{area} on the parameter $\alpha$ entering Eq.~\eqref{AB}. 
At this point, we do not apply the IR/UV cutoff, unless $\alpha \leq 0$, in which case we set $k_{UV}=1$. The results have been obtained with the use of $1024^3$ lattice. It is clear from Fig.~\ref{scaling} that the scaling is reached for $\alpha \lesssim 2$, i.e., the area parameter 
$\xi$ approaches an approximately constant value. 
At a fixed late time, the parameter $\xi$ monotonously grows as the parameter $\alpha$ decreases, see Fig.~\ref{alpha}. 
The growth continues until the value $\xi \approx 1.2$ is reached, at which point it saturates. Note that the final time of simulations $\tau_f$ has been slightly extended beyond $\tau_f \approx 22$ (see the end of the previous section) till $\tau_f \approx 25$. We deem this difference inessential; furthermore the data demonstrated in Fig.~\ref{alpha} has been collected at the ``safe'' time $\tau = 15$.

For $\alpha < 2$, the initial distribution of the scalar $\chi$ is dominated by the largest momenta 
available, while the contribution of IR modes is negligible. Nevertheless, dependence of the scaling parameter $\xi$ on $\alpha$ manifested in Fig.~\ref{scaling} is unlikely to be attributed to UV modes. Indeed, applying the cutoff $k_{UV} =1$ for the initial conditions with $\alpha \leq 0$ does not alter monotonous growth of $\xi$ as one decreases $\alpha$. This is particularly clear from Fig.~\ref{alpha}, and we provide more evidence of that in the present section. Rather dependence $\xi (\alpha)$ manifests significance of IR modes. Recall that the IR slope of the initial scalar distribution also depends on $\alpha$ in the monotonous fashion, see Eq.~\eqref{alphair}: by increasing $\alpha$, the IR part of the spectrum is amplified. Let us stress that the absolute power contained in IR modes is irrelevant; what matters is the IR slope of the scalar power spectrum. This is clear from Fig.~\ref{scaling}, where we demonstrate the scaling behavior in the case $\alpha=0.5$ as defined in Eq.~\eqref{AB} and with doubled values of $A(k)$ and $B(k)$.

It is not surprising by itself that the parameter $\alpha$ affects DW evolution. We can gain some intuition about this considering the model of a free massless scalar with initial conditions again given by Eqs.~\eqref{init} and~\eqref{AB}. In the expanding Universe, IR modes describing the initial scalar distribution are frozen at early times, because they are in the superhozion regime, while the subhorizon modes are decreasing inversely proportional to the scale factor, $\propto 1/a$. Therefore, IR modes may become relevant even if they are initially negligible. 
Whether it happens or not depends on the value of $\alpha$. Namely, for large negative $\alpha$, IR modes in the initial scalar distribution are strongly suppressed, see Eq.~\eqref{alphair}, and the cosmological expansion is not sufficient to bring them to life. This explains why the scaling parameter $\xi$ eventually reaches saturation. 
Let us stress though that this heuristic picture assuming a free scalar field completely neglects mode mixing, which is the inherent characteristic of the nonlinear system we are interested in. The only goal of the above argument is to demonstrate how cosmological evolution amplifies the relevance of IR modes\footnote{This picture can be checked by the study of DWs in the Minkowski spacetime. In fact, such simulations have been performed in the case of so-called melting DWs with a time-varying tension. In the RD Universe, melting DWs are equivalent to standard constant tension walls in the flat spacetime~\cite{Babichev:2021uvl}. In the cases of vacuum and thermal initial conditions discussed in Ref.~\cite{Dankovsky:2024ipq} we obtained very close values of the area parameter $\xi$ in the scaling regime, see the top panel in Fig.~5 there. This is explained by the fact that IR modes, which are initially suppressed in this case, will remain so at all the times, as there is no division into superhorizon and subhorizon modes.}.

For $\alpha \geq 2$ we have a qualitatively different picture, because in that case the initial scalar field configuration is dominated by IR modes according to Eq.~\eqref{alphair}. 
The initial correlation length set by the box comoving length $L$ is large in this case (relative to the initial Hubble radius), which suggests that the DW network is rather sparse at early times and hence it is characterized by the 
smaller scaling parameter $\xi$ compared to the case $\alpha <2$.  
On the other hand, the horizon radius grows faster than the physical box length $a L$, and the system tends to reach the scaling behavior, which is manifested in the growth of the area parameter $\xi$.
Physically, this picture can be explained as follows: initially there is less than one DW per horizon volume on average (exactly because DWs are sufficiently smooth), but nothing prevents DWs from entering empty Hubble patches. This is how the standard picture with one DW per horizon volume is getting restored. 
However, we do not observe the scaling regime for $\alpha \geq 2$, because it is expected to start after the Hubble radius overcomes the box size, but at these times simulations should be terminated. We do not discuss the case $\alpha \geq 2$ in what follows.

\begin{figure}[!htb]
\begin{center}
    \includegraphics[width=\textwidth]{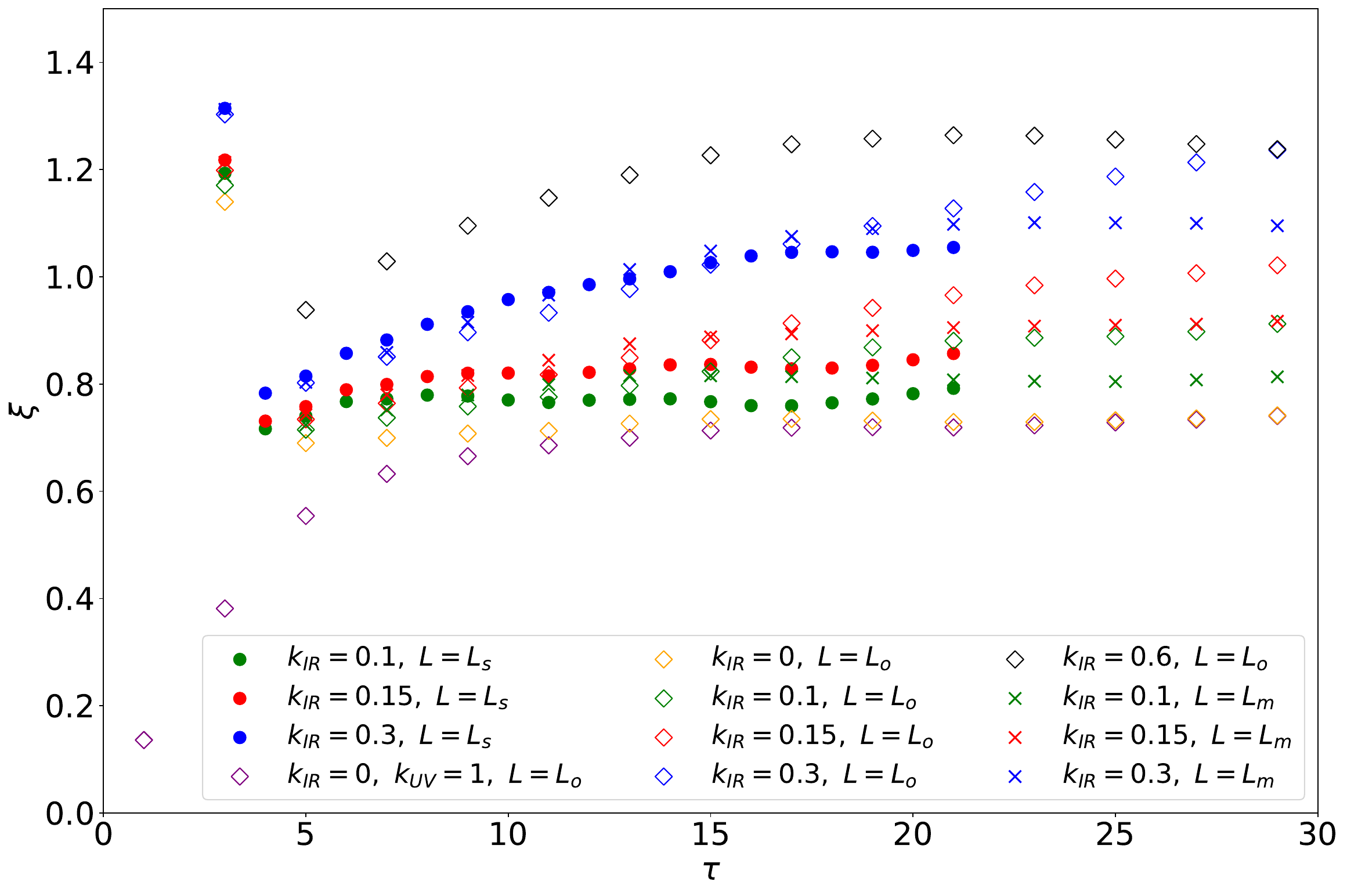} 
\end{center}
    \caption{Evolution of the area parameter $\xi$ is demonstrated for different values of the cutoff scale $k_{IR}$ parameterizing initial conditions in Eq.~\eqref{AB} with $\alpha=0.5$. No cutoff scale $k_{UV}$ is imposed, unless otherwise specified. The box size has been set either to the optimal side length $L=L_o$ given by Eq.~\eqref{optimal}, to $L_m=L_o/\sqrt{2}$, or to $L_s=L_o/2$. For $L=L_o$ and $L=L_m$, simulations have been performed with $2048^3$ lattice, and for $L=L_s$ they have been performed with $1024^3$ lattice.} \label{cutoff_alpha_05}
\end{figure}

Let us discuss in some details the impact of UV modes. They account for the large values of the scaling parameter $\xi$ observed for some of $\xi$-curves in Fig.~\ref{scaling} at early times\footnote{See Refs.~\cite{Mukhopadhyay:2024wii, Martins:2016lzc} for theoretical studies of the DW network evolution before the scaling regime.}. Indeed, the bump disappears whenever the cutoff $k_{UV}=1$ is imposed. In the case $\alpha>0$, we do not impose the cutoff $k_{UV}=1$, because there is a natural thermal cutoff $k \sim (\mbox{a few}) \times T$. Nevertheless, the latter can be high enough and one observes relatively large $\xi$ initially, which is a manifestation of sufficiently curved DWs. Note that the thermal cutoff decreases with increasing $\alpha$. This explains, e.g., why the bump is smaller in the case $\alpha=1$ compared to $\alpha=0.5$. Furthermore, imposing the cutoff $k_{UV}=1$ in the case $\alpha=0.5$, one ends up with a smooth evolution of the area parameter $\xi$ at all the times, as it is shown in Fig.~\ref{cutoff_alpha_05}. We also observe in Fig.~\ref{cutoff_alpha_05} in the case $\alpha=0.5$ that the late time behavior for the area parameter is essentially indistinguishable in the cases with $k_{UV}=1$ and with no UV cutoff imposed. One can also compare our results 
obtained in the case $\alpha =1$ with those of Ref.~\cite{first}, where the same case has been studied but for the larger ``temperature'' $T'=T_U=5T$. Results for the area parameter inferred from Fig.~\ref{alpha} and Ref.~\cite{first} are in excellent agreement with each other, despite grossly different UV properties of initial conditions.   

\begin{figure}[!htb]
\begin{center}
    \includegraphics[width=\textwidth]{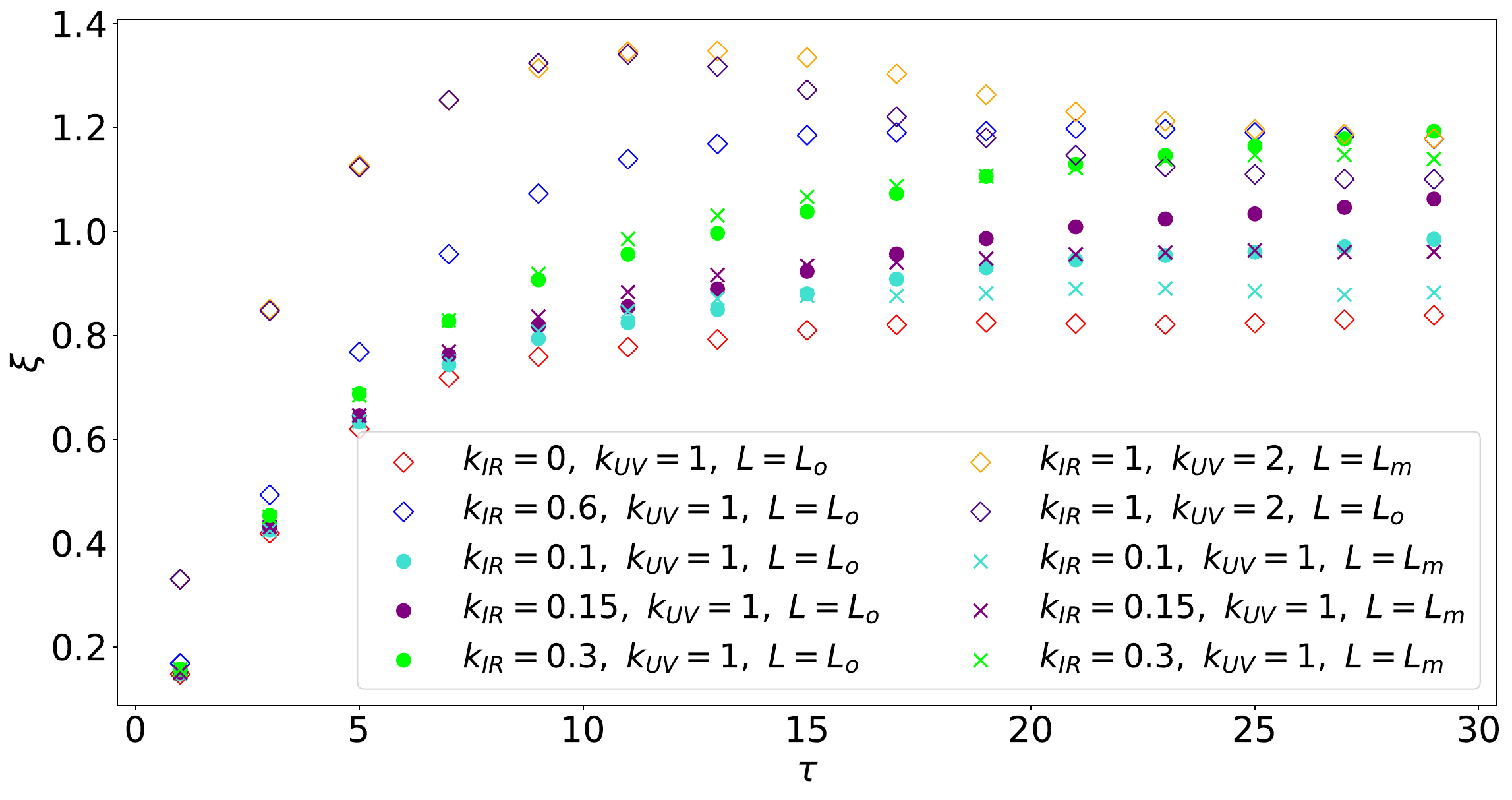} 
\end{center}
    \caption{Evolution of the area parameter $\xi$ is demonstrated for various choices of the cutoff scales $k_{IR}$ and $k_{UV}$ parameterizing initial conditions in Eq.~\eqref{AB} with $\alpha=0$. Simulations have been performed with $2048^3$ lattice. The box size has been set to $L=L_o$, given by Eq.~\eqref{optimal}, or $L_m=L_o/\sqrt{2}$.} \label{cutoff_alpha_0}
\end{figure}

To summarize, the presence of high momentum modes in the initial conditions can impact the early behavior of DWs, 
but they are fairly irrelevant at late times and hence 
for any phenomenological implications of DWs typically inferred from their late time dynamics. IR modes are qualitatively different in this regard, as they affect the value of the area parameter in the scaling regime, 
and therefore they deserve a more elaborate investigation.

{\it Is the impact of IR modes physical?} While the previous results show that IR modes in the initial scalar spectrum have a strong impact on DW evolution, it is unclear if the effect on the area parameter $\xi$ observed is physical or not. Before we delve into the results of numerical simulations, let us give two heuristic arguments in support of non-physical origin of the effect. First, it appears rather unusual that DW evolution is so selective: it remembers about IR modes, but the memory of UV modes is erased. Second, long wavelength perturbations in the initial scalar distribution serve as natural mediators between the boundary, to which IR modes are particularly sensitive, and local DW dynamics. The boundary effects are propagating through the combination of cosmological evolution of IR modes discussed in the beginning of this section and mode coupling because of nonlinearities inherent in the system.

To address this question we remove some IR modes from the initial 
scalar field configuration by imposing a non-zero value of $k_{IR}>k_{min}$. If the effect of IR modes is physical, the result should not depend on the lattice parameter $L$ and hence on $k_{min}=2\pi/L$. It is crucial here that all the model constants as well as parameters describing initial conditions including $k_{IR} \neq 2\pi/L$ are taken to be independent of $L$. Furthermore, one expects to observe a sufficiently smooth dependence on $k_{IR}$ for a fixed $L$. At this point, we have to deal with more delicate information to be extracted from the data, 
and thus we increase the resolution of simulations by switching to the $2048^3$ lattice, unless otherwise specified. We accept three choices of the box length $L$, i.e., $L_o$ corresponding to the optimal size of $2048^3$ lattice as defined in Eq.~\eqref{optimal}, $L_m=L_o/\sqrt{2}$, and the smallest one $L_s=L_o/2$. Only in the latter case we assume the grid number 
$N=1024$ and take the final time of simulations $\tau_f \approx 22$, at which point we have approximately $10$ Hubble patches in the lattice box. In both cases $L=L_o$ and $L=L_m$ we take $\tau_f \approx 30$. For $L=L_o$ one has about $30$ Hubble patches in the lattice box at $\tau_f \approx 30$, while the DW width just becomes equal to the lattice spacing, as it has been discussed in the end of sec.~\ref{sec:setting}. Switching to $L=L_m$, one decreases the lattice spacing, but reduces the number of Hubble patches at the end of simulations. However, this reduction is not dramatic at all, and one still has approximately $10$ Hubble patches at $\tau_f \approx 30$.    

We observe in Figs.~\ref{cutoff_alpha_05} and~\ref{cutoff_alpha_0} the overall trend that the scaling value of the area parameter $\xi$ in the late time regime increases upon i) eliminating a part of IR modes from the initial scalar spectrum and ii) enlarging the box size $L$. While the former can be caused by both physical origins and numerical artifacts, dependence on the lattice parameter $L$ suggests a non-physical effect. Note that this effect is unrelated to the issues with 
a small number of Hubble patches in a lattice box or a finite lattice spacing (see also below), of which we have already taken care, as it is described in the paragraph above. We argue that this effect is likely to be due to sensitivity of IR modes to the lattice boundary set by $L$. Beforehand, let us make two remarks. At the pre-scaling epoch, the opposite trend can be observed: the area parameter $\xi$ slightly decreases upon enlarging the box. We also notice the tendency that it takes a longer time for the scaling regime to be reached with larger box sizes/$k_{IR}$, and in some cases scaling is not attained within the time span of simulations. This suggests that the boundary effects are not only about smoothing DWs in the scaling era (so that $\xi$ decreases), 
but also stimulating the system to settle earlier to the self-similar solution. This also disfavors the proposition made in Ref.~\cite{first} that the onset of scaling is defined by the ratio of the DW width and the Hubble radius, i.e., that it starts at the time $\tau$ when $\delta_{wall} \simeq 0.05 H^{-1}$.

We make a judgement regarding potential boundary effects in the situations, when the scaling regime is actually reached. In particular, the scaling takes place in the case $k_{IR}=0.1$ and in the cases $k_{IR}=0.15$ and $k_{IR}=0.3$ for $L_s$ and $L_m$, as it can be seen in Fig.~\ref{cutoff_alpha_05}. We observe that in all these cases the area parameter indeed increases upon enlarging the box size $L$ (and hence mitigating boundary effects), while $k_{IR}$ is kept fixed. There is another way to demonstrate that it is the interplay between the IR modes in the initial conditions and the box size, which matters for evolution of $\xi$. One can see from Fig.~\ref{cutoff_alpha_05} that the behavior of the scaling parameter changes considerably, as $k_{IR}$ approaches the minimal value 
of the IR cutoff $k_{min} =2\pi/L$, which is due to a finite box size. Indeed, considering a smaller box with $L=L_m$ (see above), one has $k_{min} \approx 0.09$, and we observe rather strong changes in the parameter $\xi$ evolution in the range $k_{IR}=[0.1, 0.3]$. At the same time, 
imposing the UV cutoff has almost no impact on the scaling parameter at late times. 

The results in Figs.~\ref{cutoff_alpha_05} and~\ref{cutoff_alpha_0} suggest that the area parameter $\xi$ depends on $k_{IR}$ and $L$ through the ratio $k_{IR}/k_{min}$. In particular, we end up with very close values of the area parameter $\xi$ for approximately equal $k_{IR}/k_{min}$ (which is, e.g., the case for $k_{IR}=0.1$, $L=L_m$ and $k_{IR}=0.15$, $L=L_s$). Such a dependence on the ratio $k_{IR}/k_{min}$ most vividly demonstrates that the effect we are dealing with is due to the sensitivity of IR modes to the lattice boundary set by $L$. However, proving this dependence rigorously requires a more thorough analysis with the use of higher resolution simulations.

To exclude the possibility that the observed effect is due to variation of the lattice spacing $L/N$, we also performed simulations assuming the same $L/N$ and physical parameters, but different $L$. The result can be seen in Fig.~\ref{cutoff_alpha_05} by the comparison of the cases with $L=L_o,~N=2048$ and $L=L_s,~N=1024$; for concreteness we have chosen $k_{IR}=0.15$ in this analysis. Despite that we deal with the same $L/N$, evolution of the parameter $\xi$ is different in two cases: it is larger in the late time regime for a larger box. This demonstrates that a finite lattice spacing is unlikely to play a significant role in the observed effect.

The same tendencies observed in Fig.~\ref{cutoff_alpha_05} for $\alpha=0.5$ can be seen in Fig.~\ref{cutoff_alpha_0} corresponding to the case $\alpha=0$. Interestingly, the behavior of the area parameter in the case with $k_{IR}=1$ and $k_{UV}=2$ is closer to $\xi \approx 1.4$ at $\tau \simeq 12$ breaking the bound $\xi_{max} \approx 1.2$, but eventually bounces back to $\xi \lesssim 1.2$. Whenever the plateau is reached, the value $\xi$ does not exceed $\xi_{max} \approx 1.2$ in accordance with what we have observed before.

These arguments strongly indicate that the IR modes in the initial data affect the late time DW evolution, because they are more sensitive to periodic boundary conditions compared to UV modes. We expect that once this non-physical effect is eliminated, the self-similar evolution of DWs does not exhibit a dependence on the initial conditions. The way to get rid of the unphysical dependence on the boundary is to suppress IR modes in the initial scalar spectrum, in which case one approaches $\xi \approx 1.2$.
Assuming that physically any memory of initial conditions is erased during the self-similar evolution, 
one is tempted to consider $\xi \approx 1.2$ as a genuine value of the area parameter attained in the scaling regime. So, it is likely that the value $\xi \approx 1.2$ is eventually reached in all of the simulations, once the issue with the lattice boundary is eliminated. This is the conjecture, which best summarizes the results in Figs.~\ref{alpha},~\ref{cutoff_alpha_05}, and~\ref{cutoff_alpha_0} as well as the discussion in the present section.

\section{Annihilation time of biased DWs} 
\label{sec:ann}

In this section and in sec.~\ref{sec:gw} we investigate, how suppressing/eliminating IR modes in the initial scalar field distribution
affects DW phenomenology. We mainly focus on two choices of initial conditions, i.e., $\alpha=-3, k_{IR}=0, k_{UV}=1$ and 
$\alpha=0, k_{IR}=1, k_{UV}=2$. The former choice is dictated by considerations that the area parameter $\xi$ rather quickly enters 
the scaling regime and that its late time value is close to the maximal one $\xi_{max} \approx 1.2$, see Fig.~\ref{scaling} and Fig.~\ref{alpha}. According to the discussion above, this indicates that evolution of DWs is largely detached from periodic boundary conditions. The latter choice is picked for comparison: 
in this case, evolution of DWs proceeds in a rather distinct way according to Fig.\,\ref{cutoff_alpha_0}, 
at least at sufficiently early times. 

We switch on the $\epsilon$-term in Eq.~\eqref{eqdim}, which is responsible for a slight explicit breaking 
of $Z_2$-symmetry. Recall that this is the ingredient, which is commonly introduced to make the DW network 
unstable, so that it annihilates at some conformal time $\tau_{ann}$. Note that DWs with the tension $\sigma_{wall} \gtrsim (1~\mbox{MeV})^3$ should be destroyed, otherwise they spoil cosmology~\cite{Zeldovich:1974uw, Lazanu:2015fua}. We consider the symmetry breaking potential of the form
${\cal L}_{breaking}=\epsilon \chi^3$. Introducing such a term with $\epsilon \ll \lambda v$ induces a potential bias 
\begin{equation}
\label{b}
V_{bias} \approx 2 \epsilon v^3 \; .
\end{equation}
With the potential bias introduced, the whole space is dissected into the regions with true and false vacua. Initially, both vacua occupy approximately the same volume. Eventually, regions with true vacuum take more space, while the pockets, containing false vacuum, shrink, and the DW network vanishes. One can parameterize the decay of the false vacuum fraction as follows, cf. Refs.~\cite{Kitajima:2023kzu, Hindmarsh:1996xv, Larsson:1996sp, Correia:2014kqa, Pujolas:2022qvs, Ferreira:2024eru}: 
\begin{equation}
\label{fvf}
{\cal F}_{fv} =\frac{1}{2} \cdot \mbox{exp} \left[-\left(\frac{\tau}{\tau_{ann}} \right)^p \right] \; ,
\end{equation}
where $\tau_{ann}$ is the conformal annihilation time, and the parameter $p$ controls intensity of the annihilation process. Note that Eq.~\eqref{fvf} gives
\begin{equation}
{\cal F}_{fv} (\tau_{ann}) =\frac{1}{2e} \; ,
\end{equation}
which can be put forward as the definition of the annihilation time $\tau_{ann}$. Notably, it is independent of the parameter $p$.

Naively, one expects that the DW annihilation takes place at the time, when $\sigma_{wall} H (\tau_{ann}) \sim V_{bias}$, 
from which one extracts the following functional relation $\tau_{ann} \propto \epsilon^{-1/2}$. Physically this estimate assumes that the annihilation occurs once infinitely thin DWs have gained a considerable acceleration caused by the vacuum pressure~\cite{Garriga:1991ts}. Surprisingly, a different 
functional relation has been found in Ref.~\cite{bias}, i.e., $\tau_{ann} \propto \epsilon^{-1/3}$. 
Furthermore, we have found that the same functional relation follows from analysis of the simulation data obtained in Ref.~\cite{Ferreira:2024eru}.

\begin{figure}[!htb]
\begin{center}
    \includegraphics[width=\textwidth]{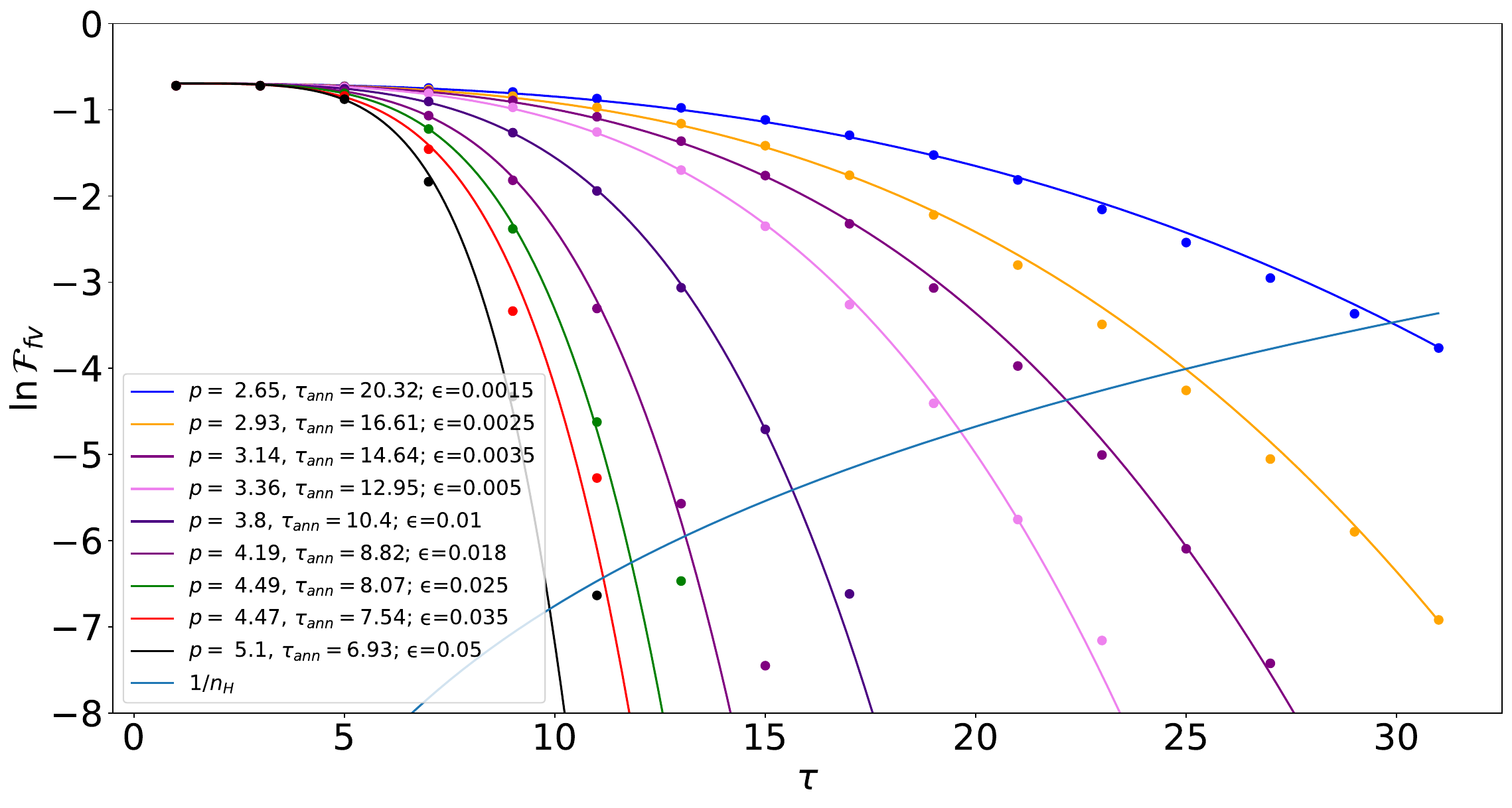} 
\end{center}
    \caption{Evolution of false vacuum fraction ${\cal F}_{fv}$ under the influence of the potential bias is demonstrated in the range $\epsilon \in [0.0015, 0.05]$. The parameterization~\eqref{fvf} has been used. Simulations have been carried out with $2048^3$ lattice.} \label{fvf12}
\end{figure}

\begin{figure}[!htb]
\begin{center}
    \includegraphics[width=\textwidth]{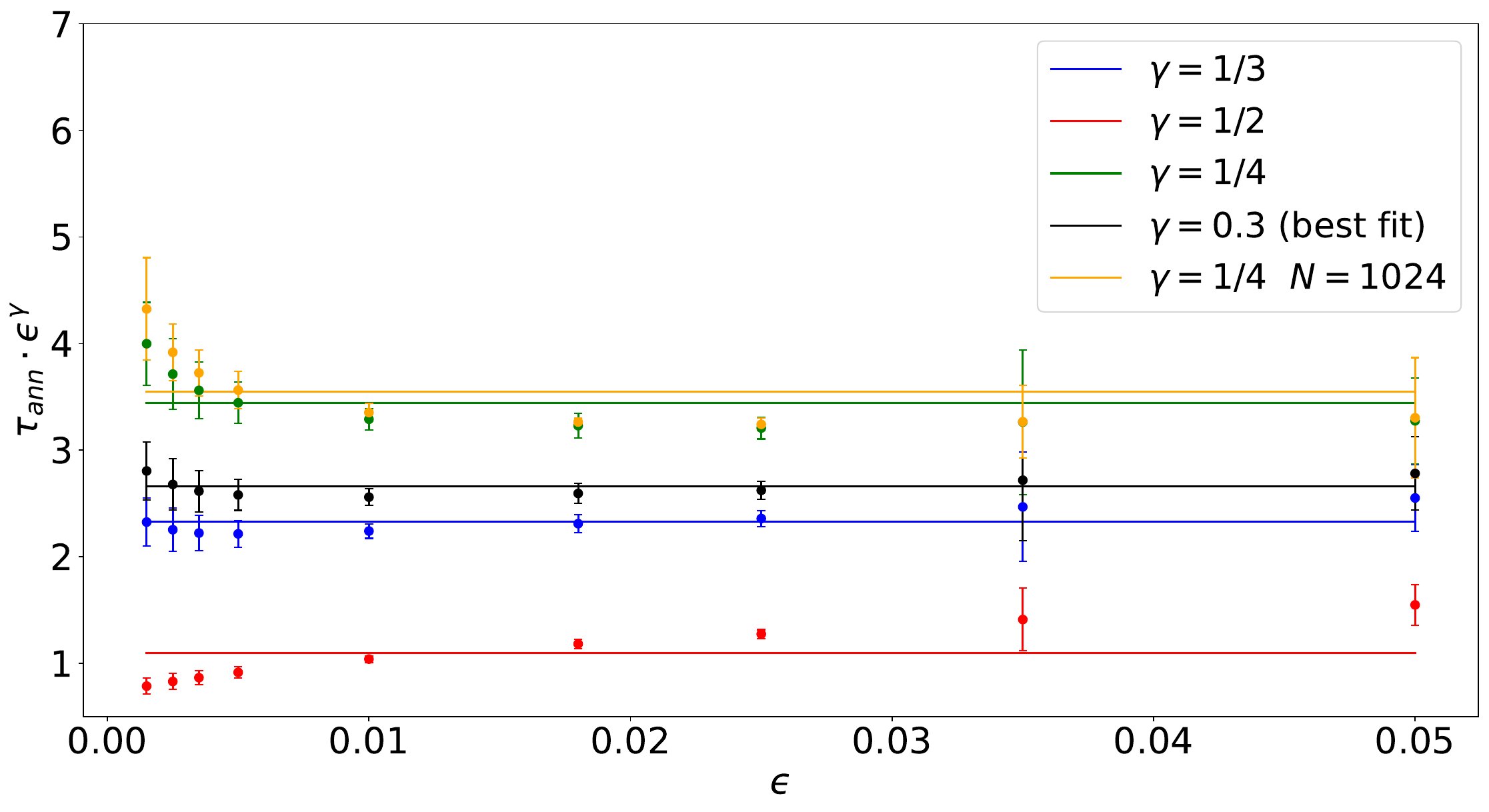} 
    \includegraphics[width=0.495\textwidth]{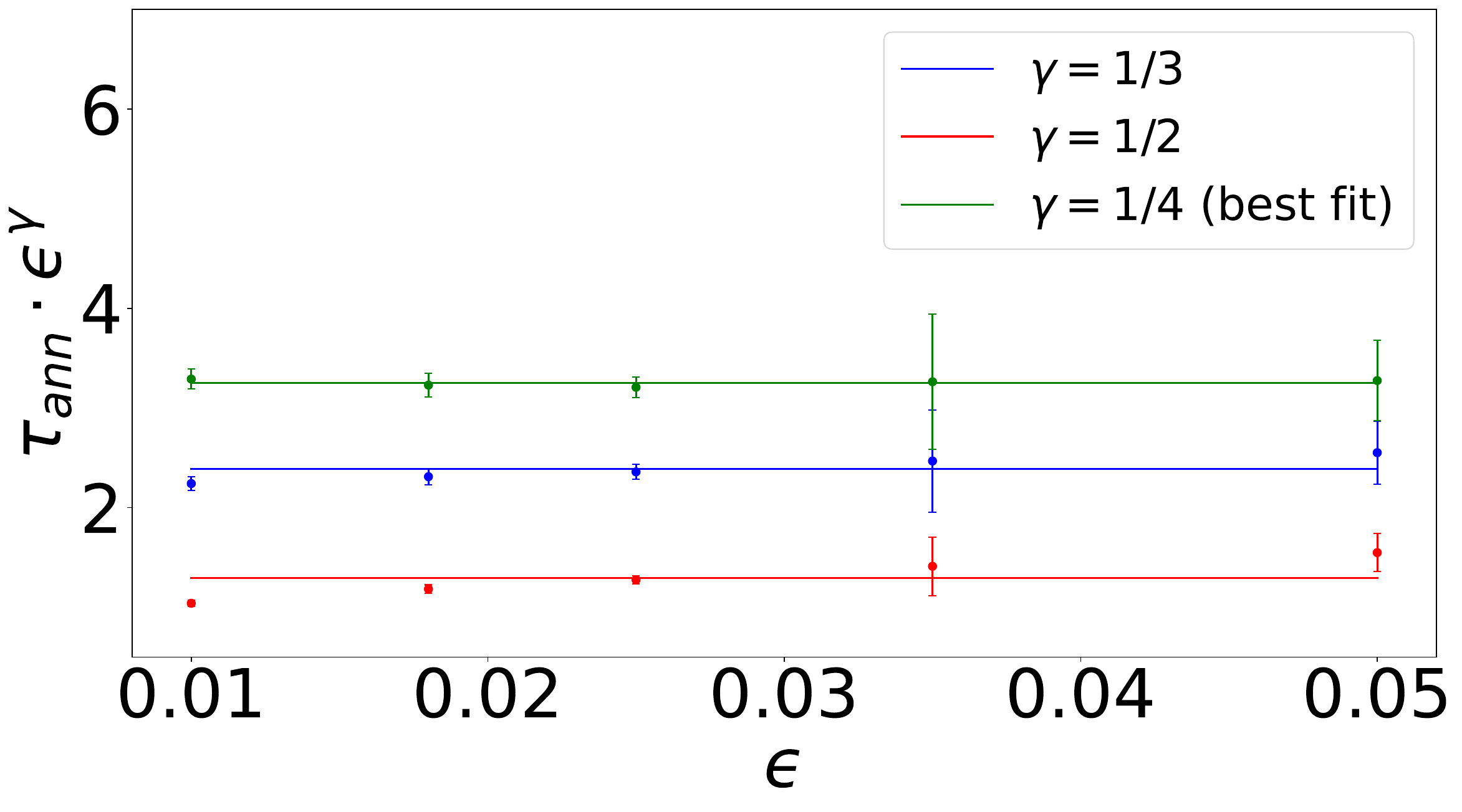}
     \includegraphics[width=0.495\textwidth]{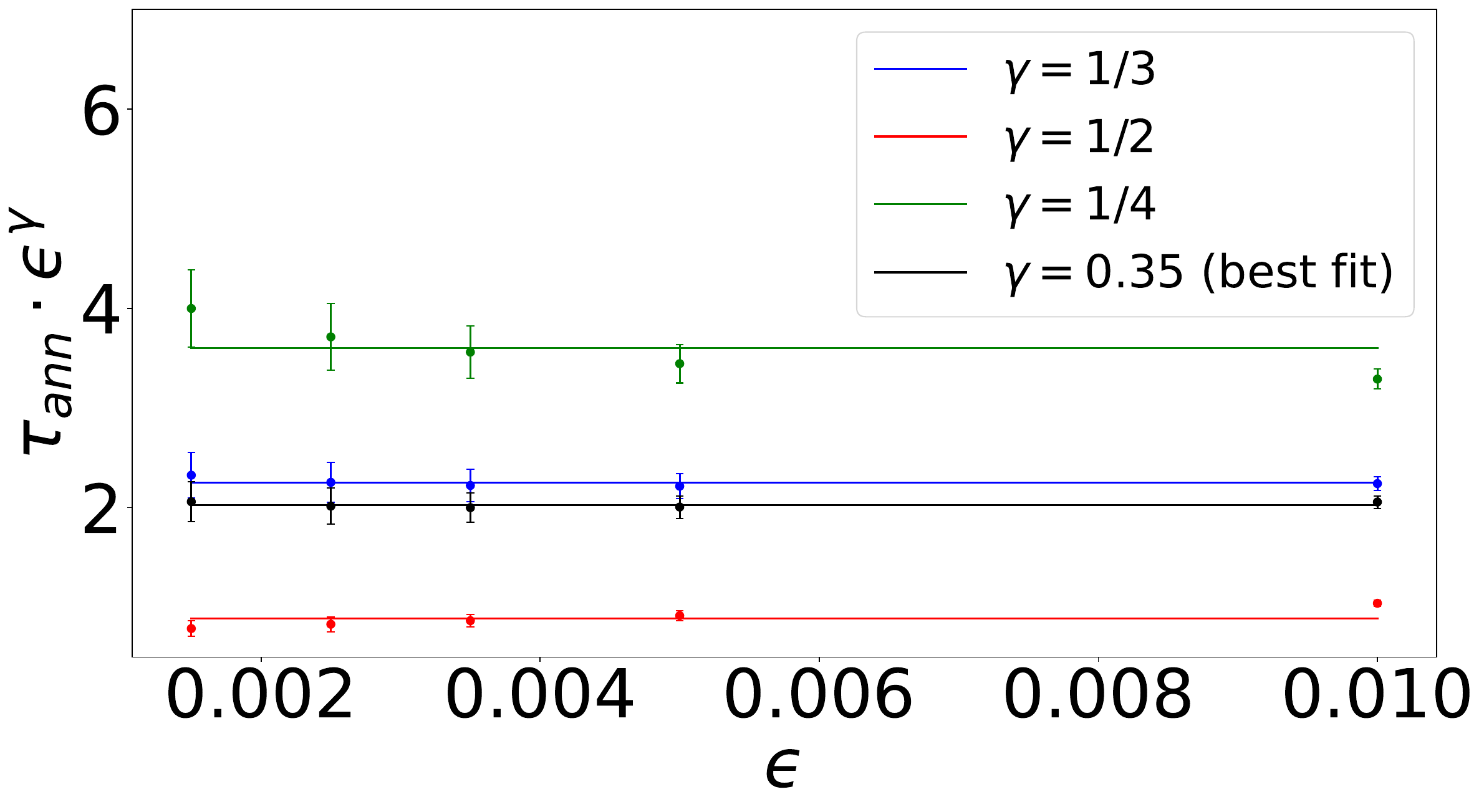}
\end{center}
    \caption{Dependence of the quantity $\tau_{ann} \epsilon^{\gamma}$ on the bias parameter $\epsilon$ is demonstrated for $\gamma=1/2, 1/3, 1/4$. The power-law fit assuming $\tau_{ann} \epsilon^{\gamma}=C$, where $C$ is a constant, is shown for the same values of $\gamma$ as well as for the best fit value of $\gamma$. The initial data with $k_{IR}=0$, $k_{UV}=1$, and $\alpha=-3$ in Eq.~\eqref{AB} have been assumed. Unless otherwise specified, simulations have been performed assuming $2048^3$ lattice box in the range of values of $\epsilon \in [0.0015, 0.05]$ (top panel); $\epsilon \in [0.0015, 0.01]$ (bottom right panel); $\epsilon \in [0.01, 0.05]$ (bottom left panel).}  \label{tau_ann_ext}
\end{figure}

It is interesting to check, how this result is affected upon removing IR modes in the initial data. The results of lattice simulations are shown in Figs.~\ref{fvf12},~\ref{tau_ann_ext} 
and in Table~\ref{table}. 
\begin{table}[htb]
    \centering
    \begin{tabular}{|c|c|c|c|c|c|c|c|c|c|}
    \hline
        $\epsilon $ &  Lattice  & $\alpha$ &$k_{UV}$ & $k_{IR}$ & $\chi^2$, $\gamma=1/2$ & $\chi^2$, $\gamma=1/3$ & $\chi^2$, $\gamma=1/4$ & $\gamma$, BF \\
        \hline 
          $[0.01, 0.05]$  & $2048^3$ & $-3$ & $1$ & $0$ & $17.9$ & $1.44$ & $0.09$  & $0.25$  \\
          \hline
              $[0.01, 0.05]$  & $1024^3$ & $-3$ & $1$ & $0$ & $39.2$ & $3.53$ &  $0.39$ & $0.26$  \\
        \hline 
          $[0.0015, 0.05]$  & $2048^3$ & $-3$ & $1$ & $0$ & $10.9$ & $0.47$ & $1.75$ & $0.3$   \\
        \hline 
            $[0.0015, 0.05]$  & $1024^3$ & $-3$ & $1$ & $0$ & $18.6$ & $1.39$ &  $13.2$ &  $0.32$  \\
        \hline 
            $[0.0025, 0.05]$  & $2048^3$ & $-3$ & $1$ & $0$ & $11.4$ & $0.54$ & $0.94$ &  $0.29$ \\
            \hline 
              $[0.0025, 0.05]$  & $1024^3$ & $-3$ & $1$ & $0$ & $17.6$ & $1.02$ & $7.0$ &  $0.31$ \\
            \hline 
                $[0.01, 0.05]$  & $1024^3$ & $0$ & $2$ & $1$ & $12$ & $1.1$ & $0.06$ & $0.25$  \\
               \hline 
             $[0.0025, 0.05]$  & $1024^3$ & $0$ & $2$ & $1$ & $26$ & $1.1$ & $0.96$ & $0.28$  \\    
             \hline 
                $[0.01, 0.05]$  & $1024^3$ & $1$ & - & $0$ & $0.43$ & $0.04$ & $0.3$ & $0.36$  \\   
            \hline
             $[0.01, 0.05]^*$  & $1024^3$ & $0$ & $1$ & $0$ &$23.1$ & $0.29$ & $7.54$ & $0.34$   \\
             \hline
                 $[0.01, 0.05]^*$  & $1024^3$ & $0$ & $0.3$ & $0$ &$2.2$ & $0.05$ & $0.16$ & $0.3$   \\ 
             \hline
        $[0.01, 0.05]^*$  & $1024^3$ & $0$ & $5$ & $0$& $4.3$ & $0.7$ & $2.2$ & $0.34$  \\
        \hline
       $ [0.0035, 0.05]^*$  & $1024^3$ & $0$ &  $1$ & $0$ &$8.6$ & $1.1$ & $4.80$ & $0.38$   \\ 
       \hline
         $[0.0035, 0.05]^*$  & $2048^3$ & $0$ & $1$ & $0$ &$42.2$ & $1.9$ & $12.4$ & $0.37$  \\
         \hline
    $[0.0009, 0.0016]^{**}$  & $3240^3$  & $0$ & $2\sqrt{2}$ & $0$ &$5.0$ & $0.05$ & $0.73$ & $0.32$  \\ 
    \hline
        $[0.0008, 0.0016]^{**}$  & $3240^3$ & $0$ & $2\sqrt{2}$ & $0$ & $3.0$ & $1.1$ & $4.95$ & $0.42$  \\ 
    \hline
    \end{tabular}
       \caption{Results of the analysis of DW annihilation caused by the potential bias are demonstrated for different initial conditions assuming the power-law dependence $\tau_{ann} \epsilon^{\gamma} =C$, where $C$ and $\gamma$ are constants. We study the cases with $\gamma$ being set to fixed values $\gamma=1/2, 1/3, 1/4$. We also show the best fit (BF) value of $\gamma$ in the last column. In the last rows we demonstrate results of the analysis performed with the data taken from $^{*}$Ref.~\cite{bias} and $^{**}$Ref.~\cite{Ferreira:2024eru}.}\label{table}
\end{table}
When analyzing the simulation data, we only consider the points fulfilling $\ln{\cal F}_{fv} \geq \frac{1}{n_H}$, where $n_H$ is the number of Hubble volumes in the box~\cite{Ferreira:2024eru}. Taking the double logarithm of Eq.~\eqref{fvf}, we obtain $p \ln \tau - p \ln \tau_{ann} = \ln \left( \ln \frac12 - \ln {\cal F}_{fv} \right) $. Similarly, we take the logarithm of $ \tau_{ann} \cdot \epsilon^{\gamma} = C$, where $\gamma$ and $C$ are constants, and obtain $ \ln \tau_{ann}= \ln C  - \gamma \ln \epsilon $. Both are linear functions with respect to $\ln \tau $ and $\ln \epsilon$, respectively. 
Now using the polyfit function from Python library NumPy, we can perform fits and obtain the parameters $\tau_{ann}$, $p$, $\gamma$ and  $C$. In Table~\ref{table}, which summarizes the behavior $\tau_{ann} (\epsilon)$ for different choices of initial conditions, we present our results as well as the results obtained with the data of Refs.~\cite{bias, Ferreira:2024eru} for comparison. One observes that the behavior $\tau_{ann} \propto \epsilon^{-1/2}$ is even more disfavored for the initial conditions with suppressed IR modes. From the plethora of datasets obtained in this work in the case of initial conditions with suppressed IR modes, we find that the exponent $\gamma$ is confined to the range: 
\begin{equation}
\label{range}
0.25 \leq \gamma \leq 0.35\; .
\end{equation}
After removing/suppressing IR modes, the behavior $\tau_{ann} \propto\epsilon^{-1/3}$ observed in Ref.~\cite{bias} still remains in a very good agreement with the data in the broad range of $\epsilon$. However, restricting to relatively large $\epsilon \in [0.01, 0.05]$ corresponding to $7 \lesssim \tau_{ann} \lesssim 10$, one notices preference for even lower $\gamma$, i.e., $\tau_{ann} \propto \epsilon^{-1/4}$. Such a restriction to large $\epsilon$ is dangerous, because the system may not have enough time to settle properly to the scaling regime in this case according to Figs.~\ref{scaling},~\ref{cutoff_alpha_05}, and~\ref{cutoff_alpha_0}. Yet we get very similar results for the cases with rather different initial conditions and quite distinct pre-scaling evolution of DWs. On the other hand, extending to very small $\epsilon$ and considering large $\tau_{ann}$ up to $\tau_{ann} \sim 20$, one starts facing issues due to a finite lattice resolution. This is particularly clear from the growing discrepancy in the annihilation time $\tau_{ann}$ obtained with $1024^3$ and $2048^3$ lattices; see the top panel of Fig.~\ref{tau_ann_ext}. Note that the preference for smaller $\gamma$ increases, as one switches from $1024^3$ to $2048^3$ box. These are the reasons why we have performed fitting of the parameter $\gamma$ assuming different ranges of $\epsilon$, as it is demonstrated in Fig.~\ref{tau_ann_ext} and in Table~\ref{table}.

In what follows, we stick to $\gamma \approx 0.3$ which is the best fit value derived from the range $\epsilon \in [0.0015, 0.05]$. Returning to the dimensionful variables, one can summarize the annihilation law for biased DWs as follows: 
\begin{equation}
\label{ann}
\frac{H_{ann}}{\sqrt{\lambda} v} \approx 0.15 \left(\frac{\epsilon}{\lambda v} \right)^{0.6} \; .
\end{equation}
Extrapolating Eq.~\eqref{ann} to tiny values of $\epsilon$, which are of main interest from the viewpoint of DW phenomenology, 
one concludes that DW annihilation occurs much earlier than previously expected. 

Let us make an important remark here. As it has been emphasized above, the dependence $\tau_{ann} (\epsilon)$ revealed in our numerical simulations is in a good agreement with the simulation data of Ref.~\cite{Ferreira:2024eru}. Yet the latter assumes the standard relation $\tau_{ann} \propto \epsilon^{-1/2}$, i.e., $\gamma=1/2$. 
The discrepancy from the case $\gamma \approx 0.3$ is not dramatic, once a small range of $\epsilon$ is concerned. However, extrapolating the behavior $\tau_{ann} (\epsilon)$ to tiny $\epsilon$ yields drastically different results in the cases $\gamma \approx 0.3$ and $\gamma = 1/2$, and this is the reason why we arrive to physically different conclusions with Ref.~\cite{Ferreira:2024eru}, e.g., regarding production of GWs (see the discussion in the next section). We would like to note that the possibility of such an extrapolation is a hypothesis and it is yet to be proven. 
Indeed, our analysis has been performed assuming values of the parameter $\epsilon$, 
for which the acceleration acquired by DWs due to the potential bias $\sim V_{bias}/\sigma_{wall}$ is non-negligible at the time $t_{ann}$. 
It is important to verify validity of Eq.~\eqref{ann} in the regime, where $t_{ann} \ll \sigma_{wall}/V_{bias}$, so that one can neglect the possible impact of DW acceleration (caused by the potential bias) and exclude it as a factor in DW annihilation.

\section{Gravitational waves}
\label{sec:gw}

\subsection{Unbiased DWs}

In this section, we study how a variation of initial conditions affects GWs. We mainly focus on the initial conditions with suppressed IR modes, for which $\xi \approx 1.2$, and consider other choices for comparison purposes. Note that the network of DWs is spatially inhomogeneous with a characteristic correlation length of the order of the inverse Hubble rate. Recall also that the energy density of DWs grows relative to that of surrounding matter. This is the reason why the DW network should be annihilated, but for the same reason one can expect observable GWs from DWs. Properties of GWs are quantified by the fractional spectral energy density:
\begin{equation}
\Omega_{gw} =\frac{1}{\rho_{tot}} \cdot \frac{d\rho_{gw}}{d\ln k} \; ,
\end{equation}
where $\rho_{tot}=3H^2 M^2_{P}$ is the total energy density of the Universe. 

\begin{figure}[!htb]
\begin{center}
    \includegraphics[width=\textwidth]{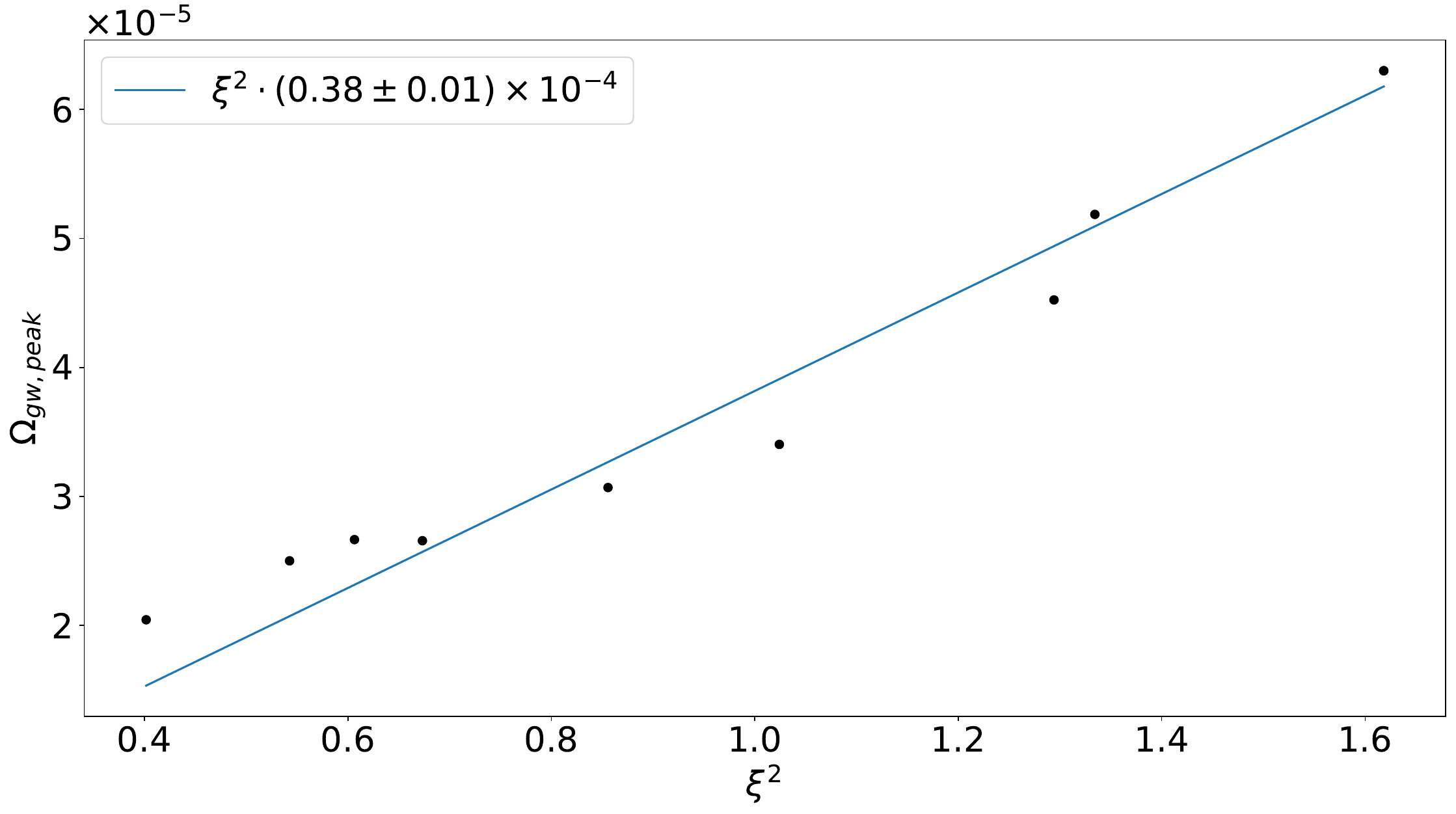} 
\end{center}
    \caption{Fractional energy density of GWs at peak $\Omega_{gw, peak}$ is shown as a function of $\xi^2$ at the conformal time $\tau=15$. 
    Initial conditions for the scalar field $\chi$ described by $\alpha=-3, -2, -1, -0.5, 0, 1/3, 0.5, 1$ in Eq.~\eqref{AB} have been assumed. The cutoff $k_{UV}=1$ (and $k_{IR}=0$) on the initial scalar spectrum has been imposed in the case $\alpha < 0$. In the case $\alpha=0$, we have considered two choices: $(k_{IR}, k_{UV})=(0,1)$ and $(k_{IR}, k_{UV})=(1,2)$. In all the other cases no IR or UV cutoff has been set.} \label{gw}
\end{figure}

We numerically evaluate GW spectra using the code CosmoLattice. In Fig.~\ref{gw} showing GW spectral energy density at peak, we allow for the lattice with $N^3=1024^2$ grid points. In all the other cases, we choose the $2048^3$ lattice with the comoving box size $L$ satisfying the condition~\eqref{optimal}. The procedure of obtaining GW spectra devised in Refs.~\cite{first, bias} is as follows. Unlike in the previous analyses where we have fixed $\kappa'/\kappa=\pi/2$ in Eq.~\eqref{optimal}, now we allow for a variation of $\kappa'/\kappa$. Namely, when computing GW spectra for any fixed initial conditions we consider two choices: $\kappa'/\kappa=2\pi$ and $\kappa'/\kappa=\pi/6$. With the latter choice, one significantly reduces $L$ (compared to the case $\kappa'/\kappa=\pi/2$) and hence the lattice spacing, thus getting a better resolution in the large momentum regime. This is crucial, because otherwise the issue with a lattice spacing emerges too early in simulations strongly compromising the UV part of the spectrum. See the discussion in the end of this subsection. However, the improvement in the UV comes at the price of spoiling IR properties of the spectrum. We partially compensate for this by taking $\kappa'/\kappa=2\pi$. In this case one increases the box size allowing to probe the spectrum in the low momentum regime, at the price of neglecting its UV part. We eliminate the points in the GW spectrum satisfying $k/a <2\pi H$ in the case $\kappa'/\kappa=2\pi$ and the points satisfying $k/a  >2\pi H $ in the case $\kappa'/\kappa=\pi/6$. Finally, we combine remaining parts of the GW spectrum for two $\kappa'/\kappa$.

First, we discuss, how initial conditions influence the lattice predictions for the properties of GWs in the peak region. 
Following Ref.~\cite{Hiramatsu:2013qaa}, we write $\Omega_{gw, peak} (\tau)$ as follows, 
\begin{equation}
\label{omega}
\Omega_{gw, peak} (\tau) =\frac{\tilde{\epsilon}_{gw} \xi^2  \sigma^2_{wall}}{24 \pi H^2 (\tau) M^4_{P}} \; ,
\end{equation}
where $\tilde{\epsilon}_{gw}$ is the dimensionless constant that reflects the efficiency of GW emission. The parameterization~\eqref{omega} is motivated by the fact that in the scaling regime the energy density of GWs remains constant, $\rho_{gw} \sim \sigma^2_{wall}/(8\pi M^2_{P})$. In turn, the latter relation stems from the fact that the energy density of GWs relies on the square of the DW stress-energy tensor. Hence, one may expect that $\Omega_{gw}$ depends on the square of the wall area parameter $\xi$, 
as it is indicated in Eq.~\eqref{omega}. 
We check this qualitative statement numerically. 
To temper the statistical fluctuations inherent in numerical simulations, we chose to average $\Omega_{gw}$ over the logarithmic momenta in the range $-0.5\leq\ln(k/k_{peak})\leq 0.5$. 
The relation $\Omega_{gw} \propto \xi^2$ is indeed fulfilled, as it follows from Fig.~\ref{gw}. 
Consequently, being agnostic about initial conditions, one should assume the factor two uncertainty regarding the parameter $\xi$, 
and consequently the factor ${\cal O} (5)$ uncertainty in $\Omega_{gw, peak}$. On the other hand, the parameter $\tilde{\epsilon}_{gw}$ carries almost no dependence on $\xi$ (in the range $0.5 \lesssim \xi \lesssim 1.2$) and initial conditions. The value of $\tilde{\epsilon}_{gw}$ inferred from the comparison of Fig.~\ref{gw} and Eq.~\eqref{omega}, where we substitute values of the model constants given in Sec.~\ref{sec:model}, reads
\begin{equation}
\label{valuegw}
\tilde{\epsilon}_{gw} \approx 0.17 \; .
\end{equation}
This agrees well with Ref.~\cite{first}, but it is somewhat below the value inferred in Ref.~\cite{Hiramatsu:2013qaa}. Let us write down the expression for $\Omega_{gw, peak} (\tau)$ in terms of model constants: 
\begin{equation}
\label{gwpeakunbiased}
\Omega_{gw, peak} (\tau) \approx 1.1 \cdot 10^{-9} \cdot \left(\frac{H_i}{H(\tau)} \right)^2 \cdot \left(\frac{v}{6 \cdot 10^{16}~\mbox{GeV}} \right)^4 \; .
\end{equation}
Here we used $\tilde{\epsilon}_{gw} \approx 0.17$ and $\xi \approx 1.2$, which correspond to the initial scalar spectrum with suppressed IR modes. We obtained slightly larger $\Omega_{gw, peak} (\tau)$ compared to Ref.~\cite{first} dealing with vacuum and thermal initial conditions, which is well explained by the larger value of $\xi$. 

\begin{figure}[!htb]
\begin{center}
\includegraphics[width=\textwidth]{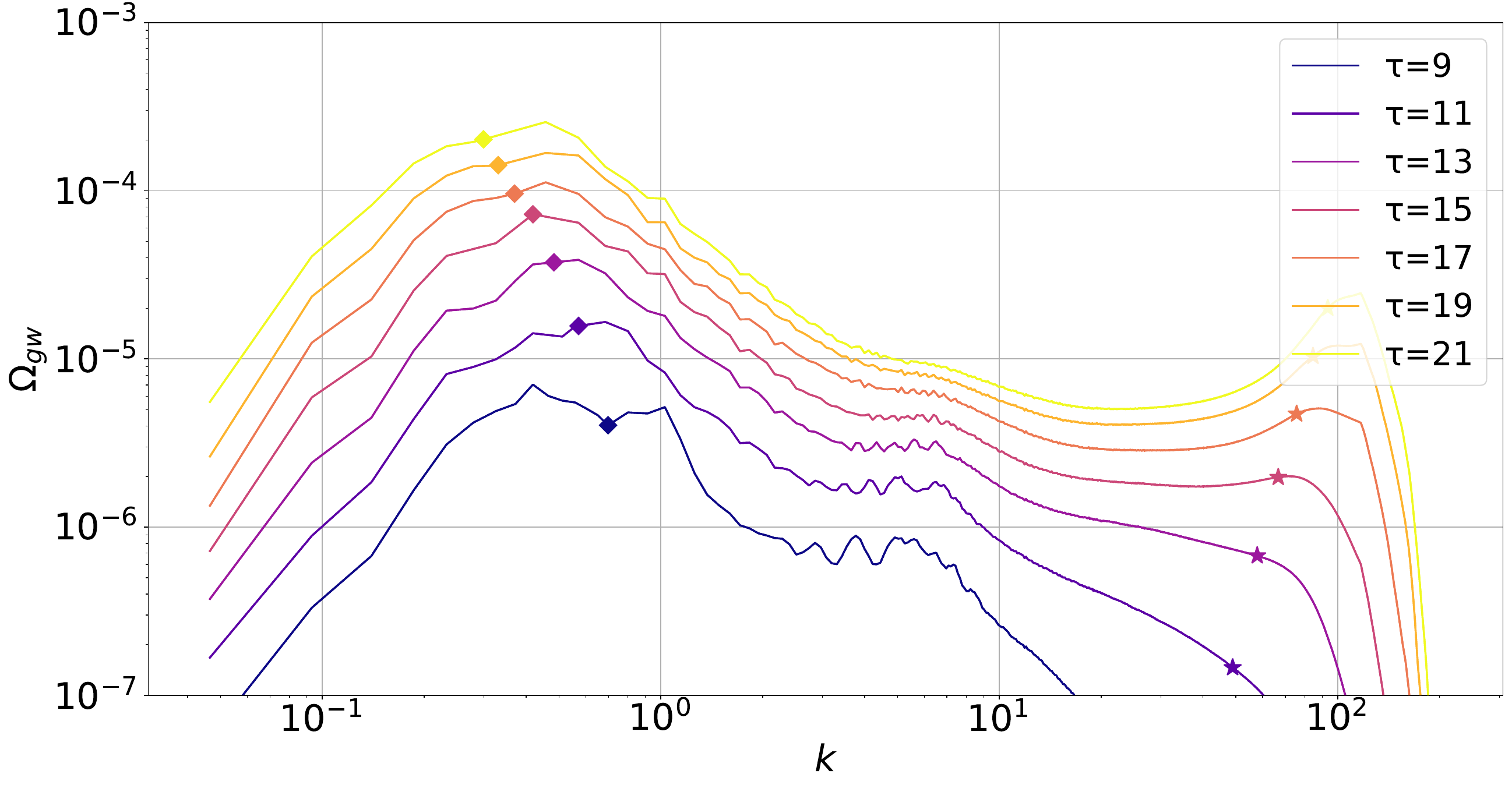} 
\end{center}
    \caption{Spectra of GWs emitted by unbiased DWs. Simulations have been executed with $2048^3$ lattice assuming initial conditions with $\alpha=-3$, $k_{IR}=0$, $k_{UV}=1$ in Eq.~\eqref{AB}. Stars and diamonds indicate the positions of the DW width and Hubble parameter in conformal variables, i.e., $k=2\pi a/\delta_{wall}$ and $k=2\pi a H$, at the given cosmic times.} \label{spectrum}
\end{figure}

\begin{figure}[!htb]
\begin{center}
    \includegraphics[width=\textwidth]{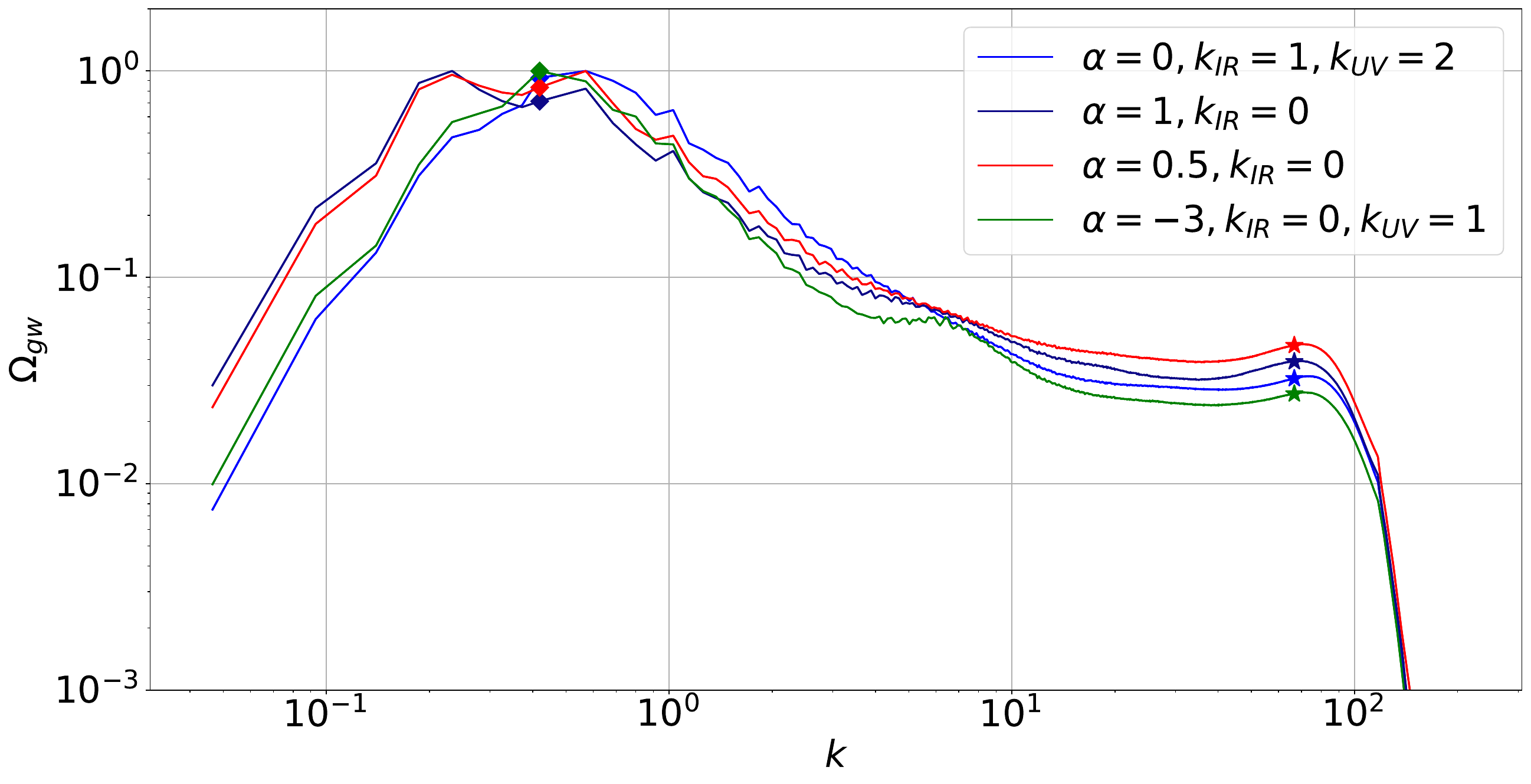} 
\end{center}
    \caption{Comparison of GW spectra from unbiased DWs at the conformal time $\tau=15$. The spectra, which are normalized to unity at peak, have been obtained assuming different initial conditions with the use of $2048^3$ lattice. Stars and diamonds indicate the positions of the DW width and Hubble parameter in conformal variables, i.e., $k=2\pi a/\delta_{wall}$ and $k=2\pi a H$,  at the given cosmic time.} \label{comparison_15}
\end{figure}

The characteristic frequency of GWs emitted about the time $\tau$ is of the order of the Hubble rate $H(\tau)$, and this estimate remains intact upon eliminating IR modes 
in the initial data, see Fig.~\ref{spectrum}. Namely, one has for the present day peak frequency $f_{peak} \simeq H(\tau)~\cdot ~a (\tau)/a_0$, where $a_0$ is the current scale factor.  
The IR slope of GW spectrum $f \ll f_{peak}$, which is very close to 
\[
\Omega_{gw} \propto k^2
\]
at all the times of simulations, at least for one decade of frequencies below the peak. This appears to be in tension with generic causality considerations~\cite{Caprini:2009fx}, which give $\Omega_{gw} \propto k^3$ for $k \rightarrow 0$, see also Refs.~\cite{Cai:2019cdl,Hook:2020phx}. Note, however, that we are unable to take the limit $k \rightarrow 0$ in our simulations, and this may explain why we observe the departure from the behavior $\Omega_{gw} \propto k^3$: in the region close to the peak the slope is gentle. 

As one can see from Fig.~\ref{comparison_15}, GW spectral shapes look qualitatively similar for different initial conditions. However, the peak frequency is slightly shifted to the left in the case of initial conditions with enhanced IR modes. This agrees well with Ref.~\cite{first} dealing with vacuum and thermal initial conditions, where the peak frequency $f_{peak} \simeq 0.7 H(\tau) \cdot a(\tau)/a_0$ has been observed. Furthermore, the near-peak region of GW spectra
is somewhat broader in the situation with larger IR modes.  
There is also some small difference in the close-to-maximum UV range: we observe a slightly sharper decrease of the power 
when IR modes are removed or suppressed.

As in Refs.~\cite{first, bias} one observes excess of GW power in the far UV range of the spectrum
(compared to the aforementioned power-law decrease). This power excess is clearly unrelated to the issue with IR modes 
in the initial conditions, and its proper analysis is out of the scope of this work. However, let us make a few comments on the far UV region, since this is arguably the most intriguing part of GW spectrum. In fact, this region can be split in several smaller regions. There is a wiggly region seen in Fig.~\ref{spectrum} around $k \sim 7$ at $\tau \sim 10$, but the wiggles vanish or become unnoticeable at sufficiently late times. Furthermore, 
these wiggles are not pronounced in the case with $\alpha=0$, $k_{IR}=1$, $k_{UV}=2$. At larger momenta, one observes a plateau; remarkably it appears independently of initial conditions. The onset of the plateau takes place at the momentum $k/a \sim 4\pi \sqrt{H/\delta_{wall}}$, which is 
the interplay between the Hubble rate and the DW width\footnote{This explains, in particular, why no such a specific far UV region is observed in the case of melting DWs~\cite{Dankovsky:2024ipq}, which effectively evolve in the Minkowski spacetime.}. 
At even larger momenta, one observes a growing peak. Generically, appearance of a peak at late times in the deep UV 
is not surprising, and it can be attributed to a finite lattice resolution~\cite{first}. However, 
the UV peak seen in Fig.~\ref{spectrum} may have a different origin, as it is correlated with the inverse DW width. 
To the right of this UV peak one observes an unphysical feature in the GW spectrum becoming more pronounced with time and developing
into the above mentioned artificial peak, which will eventually absorb the physical UV peak. This precludes a proper analysis of the latter. In the future it is important to mitigate this obstacle and develop a proper understanding of the physical UV peak, its origin and dynamics.

\subsection{Biased DWs}
Now we turn on a slight explicit breaking of $Z_2$-symmetry and consider spectra of GWs from biased DWs. 
The first studies of this kind have been carried out in Refs.~\cite{Kitajima:2023cek, Cyr:2025nzf, Notari:2025kqq, bias}. In particular, Refs.~\cite{Cyr:2025nzf, bias} reveal a significantly 
larger power of GWs in the UV part of the spectrum. Furthermore, Refs.~\cite{Kitajima:2023cek, Cyr:2025nzf, Notari:2025kqq, bias, Ferreira:2024eru}
agree that a significant production of GWs continues after the DW annihilation time defined through Eq.~\eqref{fvf}. This observation has led to a 
claim that the energy density of GWs is larger than what was naively expected. 
As we have discussed in Sec.~\ref{sec:ann}, however, numerical simulations indicate that the 
annihilation of DWs happens earlier than commonly thought. Therefore, the amplitude of GWs is
in fact much smaller compared to naive expectations assuming that we are dealing with tiny $V_{bias}$.

\begin{figure}[!htb]
\begin{center}
    \includegraphics[width=\textwidth]{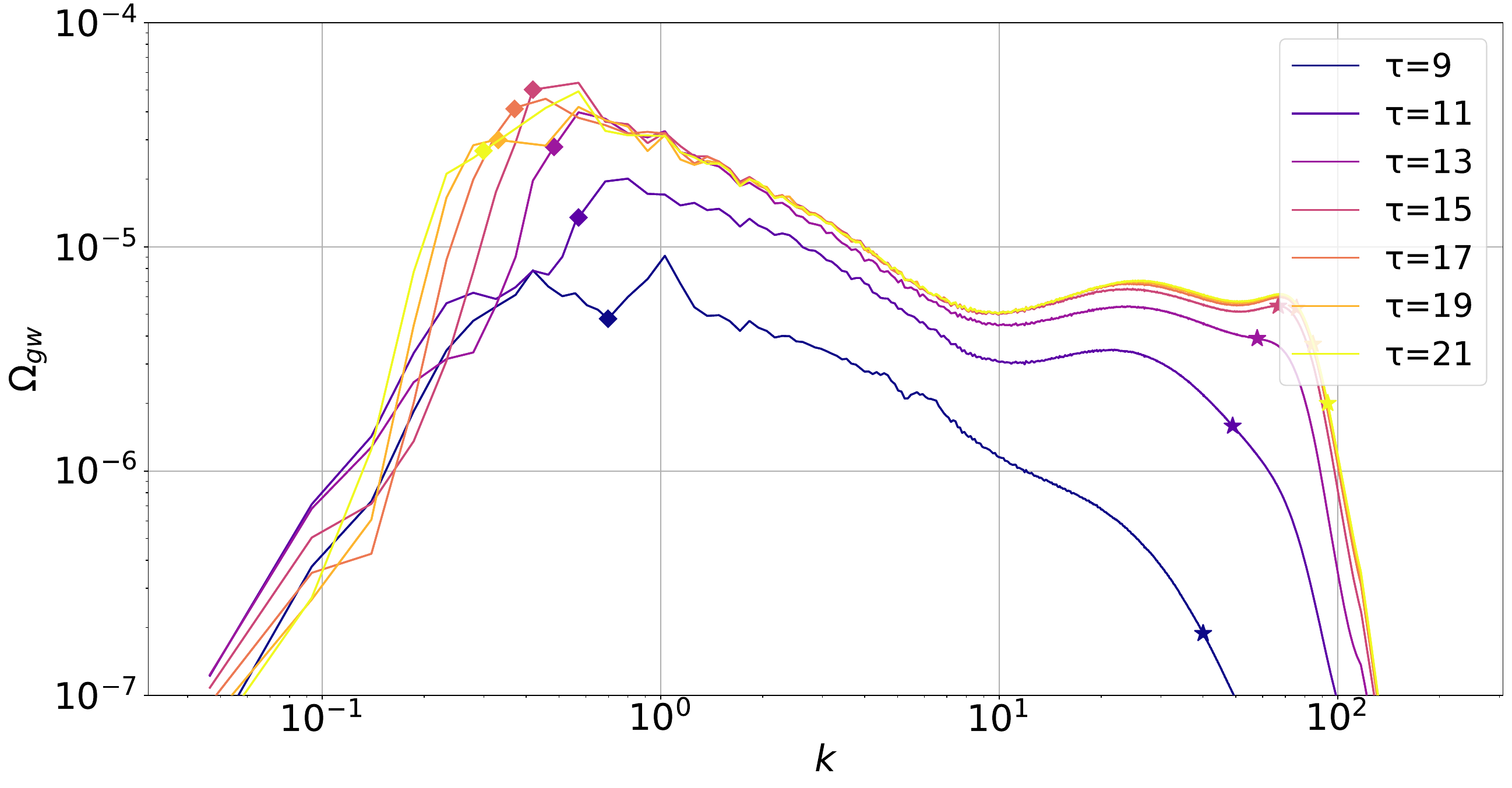} 
\end{center}
    \caption{Spectra of GWs from biased DWs at different conformal times $\tau$. The bias parameter $\epsilon$ has been 
    set to $\epsilon=0.025$. The spectra have been obtained using $2048^3$ lattice. Initial conditions with $\alpha=-3$, $k_{IR}=0$, $k_{UV}=1$ in Eq.~\eqref{AB} have been assumed. Stars and diamonds indicate the positions of the DW width and Hubble parameter in conformal variables, i.e., $k=2\pi a /\delta_{wall}$ and $k=2\pi a H$, at the given cosmic times.} \label{gw_bias_3}
\end{figure}

Our goal is to examine how the spectrum of GWs from biased DWs changes upon removing 
IR modes from the initial scalar spectrum. The result for the choice of initial conditions $\alpha=-3$, $k_{IR}=0$, $k_{UV}=1$ is 
shown in Fig.~\ref{gw_bias_3}. We set the bias parameter $\epsilon$ to a particular value $\epsilon=0.025$. For larger $\epsilon$ we might be sensitive to the initial 
DW configuration, while for smaller $\epsilon$ the time needed to produce most of GWs can be too long. 
In the case $\epsilon =0.025$ the GW spectrum gets saturated around the time $\tau \simeq 15$. Since then, the
source of GWs has been terminated, their energy density redshifts $\rho_{gw} \propto 1/a^4$, while their spectral shape remains intact. 

\begin{figure}[!htb]
\begin{center}
   \includegraphics[width=\textwidth]{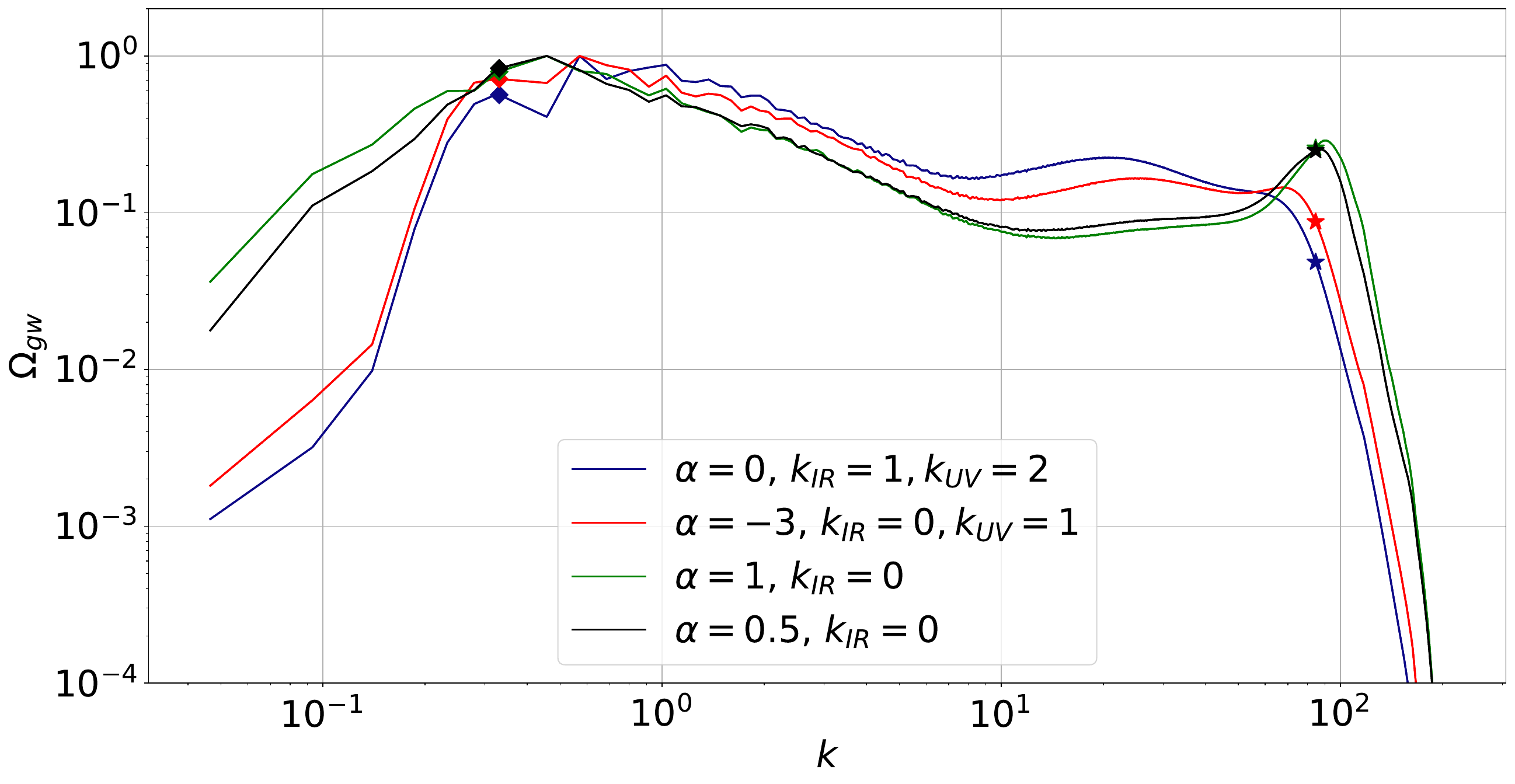} 
\end{center}
    \caption{Comparison of GW spectra normalized to unity at peak obtained from biased DWs assuming different initial conditions. The conformal time is fixed at $\tau=19$. Simulations have been performed with the use of $2048^3$ lattice. Stars and diamonds indicate the positions of the DW width and Hubble parameter in conformal variables, i.e., $k=2\pi a/\delta_{wall}$ and $k=2\pi a H$, at the given cosmic time.
    } \label{gw_bias_0_3}
\end{figure}

One can define the conformal time $\tau_{gw}$, when Eq.~\eqref{gwpeakunbiased} becomes equal to $\Omega_{gw, peak}$ at GW saturation as inferred from simulations of biased DWs. From Fig.~\ref{gw_bias_3}, we get 
\begin{equation}
\tau_{gw} \approx 1.8 \tau_{ann} \; ,
\end{equation}
which is in a good agreement with the estimate of Ref.~\cite{bias}. It means that the generation of GWs indeed takes place after the DW annihilation time, in accordance with the discussion above. Assuming the DW annihilation law~\eqref{ann}, we estimate the present day peak frequency of GWs $f_{peak} \simeq H (\tau_{gw}) \cdot (a(\tau_{gw})/a_0)$ in terms of model parameters as
\begin{equation}
\label{fpeak}
f_{peak} \simeq \frac{12~\mbox{nHz}}{\lambda^{1/20} g^{1/12}_* (T_{ann})} \cdot \left(\frac{V^{1/4}_{bias}}{ 10~\mbox{keV}} \right)^{1.2} \cdot \left(\frac{100~\mbox{TeV}}{v} \right)^{0.7}.
\end{equation}
We have ignored variation of $g_* (T)$ between the times $\tau_{ann}$ and $\tau_{gw}$. The fractional energy density of GWs at present is given by 
\begin{equation}
\label{omegapeak}
\Omega_{gw, peak} h^2_0 \simeq 1.4 \cdot 10^{-10} \cdot \lambda^{1.2} \left(\frac{v}{100~\mbox{TeV}}  \right)^{8.8} \cdot \left(\frac{10~\mbox{keV}}{V^{1/4}_{bias}} \right)^{4.8} \cdot \left(\frac{100}{g_* (T_{ann})} \right)^{1/3}. 
\end{equation}
Recall that the potential bias $V_{bias}$ and the bias parameter $\epsilon$ are related to each other by Eq.~\eqref{b}. When writing Eqs.~\eqref{fpeak} and~\eqref{omegapeak}, we have set the value of $\gamma$ describing the DW annihilation law $\tau_{ann} \propto V^{-\gamma}_{bias}$ to $\gamma \approx 0.3$, but one should keep in mind that currently $\gamma$ is subject to uncertainty, see Eq.~\eqref{range}. While not looking particularly 
dramatic, this uncertainty can strongly influence the estimates for $f_{peak}$ and $\Omega_{gw, peak}$, because one typically chooses tiny $V_{bias}$ (relative to $v^4$) for GWs to be accessible by observations. This is manifested in Eqs.~\eqref{fpeak} and~\eqref{omegapeak}, where the reference values of model constants correspond to $f_{peak}$ and $\Omega_{gw,peak}$ in the PTA range. Compared to Ref.~\cite{Ferreira:2022zzo}, our Eqs.~\eqref{fpeak} and~\eqref{omegapeak} suggest that much smaller values of $V_{bias}$, i.e., $V_{bias} \sim (10~\mbox{keV})^4$ vs. $V_{bias} \sim (10~\mbox{MeV})^4$ in Ref.~\cite{Ferreira:2022zzo}, should be considered to reproduce the observed nanohertz GW background. The reason is that different functional relations of $\tau_{ann}$ on $V_{bias}$ have been obtained in our work and supposed in Ref.~\cite{Ferreira:2022zzo}. 

\begin{table}[h]
    \centering
    \begin{tabular}{|c|c|c|c|c|c|c|c|c|}
    \hline
    $\alpha$ & $-3$ & $-3$  & $0$  & $0$\\ 
    \hline
        $\tau $ & $15$ & $19$ & $15$ & $19$   \\
     \hline
            $n_{IR}$  & $2.81 \pm 0.23$ & $2.93 \pm 0.32$  & $2.96 \pm 0.36$  & $3.09 \pm 0.34$     \\
            \hline
             $n_{UV}$  & $-0.83 \pm 0.02$  & $-0.78 \pm 0.02$  & $-0.64 \pm 0.04$  & $-0.73 \pm 0.03$   \\ 
    \hline
  \end{tabular}
       \caption{Spectral indices $n_{IR}$ and $n_{UV}$ describing power-law fits to IR and UV slopes of GW spectra from biased DWs, respectively, are demonstrated in the case of initial conditions characterized by $\alpha=-3$, $k_{IR}=0$, $k_{UV}=1$, and $\alpha=0$, $k_{IR}=1$, $k_{UV}=2$. For each choice of initial conditions, the results are shown at two different conformal times after GW production from DWs has reached saturation.}\label{powerlaw}
\end{table}

In Fig.~\ref{gw_bias_3}, we demonstrate GW spectra from biased DWs assuming the parameter $\epsilon=0.025$ in the case of initial conditions with $\alpha=-3$. We also compare the shapes of GW spectra for different $\alpha$ in Fig.~\ref{gw_bias_0_3}. We observe that the shapes 
obtained in the cases $\alpha=-3$ (with the cutoff $k_{UV}=1$ imposed) and $\alpha=0$ (with the cutoffs $k_{IR}=1$ and $k_{UV}=2$ imposed) corresponding to suppressed IR modes in the initial conditions are very similar both in the low and high GW frequency ranges. We demonstrate the spectral indices characterizing IR and UV slopes of GW spectra in Table~\ref{powerlaw}. When performing power-law fits of IR slopes, we have considered the range of momenta $k \in (k_{min}, k_{peak})$. In the case of UV slopes, we have restricted to the range of momenta $k \in (k_{peak} , k_*/\sqrt{e})$, where $k_*$ stands for the onset of the plateau. 
Note that the spectral indices describing IR tails of GW spectra should be taken with caution, because they are subject to strong uncertainties caused by a limited number of $k$-modes in the IR range. Furthermore, one can clearly see artifact oscillations plaguing the IR and near peak ranges of GW spectra. 
Such oscillations should be averaged out over GW periods when evaluating GW spectra. However due to a limited time run of simulations we have $k \tau \lesssim 1$ in the ranges of interest, so that averaging is not possible. One can see in Fig.~\ref{gw_bias_0_3}, that the GW shapes in the cases with $\alpha=-3$ and $\alpha=0$ are markedly different from the ones obtained using the initial conditions with $\alpha=0.5$ and $\alpha=1$, which are vulnerable to boundary effects. The most drastic difference is observed in the close-to-maximum low frequency range, where the GW spectra described by $\alpha=0.5$ and $\alpha=1$ exhibit a much softer decrease compared to $\alpha=-3$ and $\alpha=0$. This trend has been observed already in Fig.~\ref{comparison_15}, and it gets exacerbated for GWs from biased DWs.

\section{Discussion}
\label{sec:discussion}
We have shown that evolution of DWs as seen on a lattice depends on statistical properties of the initial conditions for the scalar field constituting DWs. This dependence comes primarily from the contribution of longer wavelength modes to the initial scalar configuration. We demonstrated, however, that the dependence on IR modes is likely to be of non-physical origin and it is due to their sensitivity to the lattice boundary, where periodic conditions are imposed. This questions reliability of lattice simulations for predictions of cosmic DW evolution. 
We observed, however, that the monotonously growing area parameter $\xi$ reaches saturation at the value $\xi \approx 1.2$ upon removing IR modes from initial conditions. Insisting that evolution of realistic cosmic DWs in the scaling regime must be independent of initial conditions, one regards $\xi \approx 1.2$ as a true value of the area parameter attained in the situation when the boundary effects are efficiently suppressed. This suggests a simple way to properly choose initial conditions for the non-ambiguous study of DW evolution: 
they should be such that the area parameter reaches $\xi \approx 1.2$ in the scaling regime.

The difference between $\xi \approx 1.2$ and the value of the area parameter commonly quoted in the literature for the case of ``vacuum'' initial conditions, i.e., $\xi \approx 0.8$~\cite{Hiramatsu:2013qaa, Ferreira:2023jbu, first}, is not dramatic. However, this motivates revisiting other aspects of DW evolution, where the impact of initial conditions can be more significant. In particular, we have revisited the annihilation of DWs under the influence of the potential bias $V_{bias}$ in the situation when IR modes in the initial conditions are removed/suppressed. We investigated the annihilation conformal time of DWs assuming the power-law dependence $\tau_{ann} \propto V^{-\gamma}_{bias}$. We strengthen the claim of Ref.~\cite{bias} that the commonly accepted value $\gamma=1/2$ is largely disfavored by the data. Compared to Ref.~\cite{bias} showing the preference for $\gamma =1/3$, we are less conclusive about the value of $\gamma$ and find that the latter is confined to the narrow range $0.25 \leq \gamma \leq 0.35$. It is possible that the preference for $\gamma \simeq 1/4$ observed in some datasets is due to elimination of IR modes, but one needs higher resolution simulations to prove this. To complement the studies of numerical simulations, it would be comforting to find a theoretically motivated model of DW annihilation, which could allow for an analytic evaluation of the exponent $\gamma$.

We have also revisited lattice evolution of GWs from DWs, biased and unbiased ones, for different choices of initial conditions. We have established on the firm footing the relation $\Omega_{gw, peak} \propto \xi^2$ in the situation with unbiased DWs, which is vividly illustrated in Fig.~\ref{gw}. In the case of initial conditions, for which the network evolution is less sensitive to boundary conditions, GW spectra shown in Figs.~\ref{comparison_15} and~\ref{gw_bias_0_3} exhibit smaller power (relative to the peak) in the close-to-maximum IR range compared to the cases where boundary effects matter. This reduction of GW power in the low frequency range is particularly pronounced for biased DWs. At the current stage of our study, the IR region of the GW spectra is still quite limited, and we would like to extend this region considerably using higher resolution simulations. With higher resolution simulations at hand, we also hope to understand better GW spectra in the high-frequency range, in particular the origin of the plateau seen in Figs.~\ref{comparison_15} and~\ref{gw_bias_0_3}. This feature can be of great importance for discriminating DWs from other sources of GW emission, and thus it is worth a thorough investigation in the future.

\section*{Acknowledgments}

Numerical simulations have been executed on the cluster of the Theoretical Division of INR RAS and on the cluster of CEICO, FZU (Phoebe). 
It is a pleasure to thank Josef Dvo\v ra\v c\'ek for the invaluable help with carrying out numerical simulations on the cluster Phoebe. D.~G. and I.~D. acknowledge the support of the scientific program of the National Center for Physics and Mathematics, section 5 ``Particle Physics and Cosmology", stage 2023-2025. I.~D. acknowledges the support by the Foundation for the Advancement of Theoretical Physics and Mathematics ``BASIS''. 
The work of E.~B. was supported by ANR grant StronG (No. ANR-22-CE31-0015-01).
The work of A.~V. was supported by the Czech Science Foundation, GA\v{C}R,
project number 24-13079S.


\end{document}